\documentclass[traditabstract,longauth]{aa}
\DeclareUnicodeCharacter{0303}{\~{n}}
\usepackage{CJK}
\usepackage{graphicx}
\usepackage{txfonts}
\usepackage{lipsum}
\usepackage{subcaption}
\usepackage{amsmath}
\usepackage{amssymb}
\usepackage{siunitx}
\usepackage{gensymb}
\usepackage{lscape}
\usepackage{placeins}
\makeatletter
\@ifundefined{linenumbers}{}{
  \let\linenumbers\relax
  \let\endlinenumbers\relax
}
\makeatother
\usepackage[colorlinks=true, linkcolor=blue, citecolor=blue, urlcolor=blue]{hyperref}
\titlerunning{AVID}
\authorrunning {Sun et al.}
\begin{document}

\title{AVID: Formation and evolution of a coalesced major merger of late-type dwarf galaxies (VCC 479) on the outskirts of the Virgo cluster}

\subtitle{}
\author{Weibin Sun\inst{1,2}
\and Hong-Xin Zhang\inst{1,2}\fnmsep\thanks{Corresponding author: hzhang18@ustc.edu.cn}
\and Rory Smith\inst{3,4}
\and Elias Brinks\inst{5}
\and Patrick C\^ot\'e\inst{6}
\and Se-Heon Oh\inst{7}
\and Zesen Lin\inst{8}
\and Alessandro Boselli\inst{9}
\and Laura Ferrarese\inst{6}
\and Fujia Li\inst{1,2}
\and Yu-Zhu Sun\inst{1,2}
\and Lijun Chen\inst{1,2}
\and Lanyue Zhang\inst{1,2}
\and Minsu Kim\inst{7}
\and Jaebeom Kim\inst{7}
\and Tie Li\inst{1,2}
\and Bojun Tao\inst{1,2}
\and Matt Taylor\inst{10}
\and Pierre-Alain Duc\inst{11}
\and Ruben S\'anchez-Janss\'en\inst{12}
\and Yinghe Zhao\inst{13,14}
\and Sanjaya Paudel\inst{15,16}
\and Eric W. Peng\inst{17}
\and Kaixiang Wang\inst{18,19}
\and Stephen Gwyn\inst{6}
\and Matteo Fossati\inst{20}
\and Jean-Charles Cuillandre\inst{21}
}

\institute{
Department of Astronomy, University of Science and Technology of China, Hefei 230026, China
\and School of Astronomy and Space Science, University of Science and Technology of China, Hefei 230026, China
\and Departamento de Fisica, Universidad Tecnica Federico Santa Maria, Avenida España 1680, Valparaíso, Chile
\and Millenium Nucleus for Galaxies (MINGAL)
\and Centre for Astrophysics Research, University of Hertfordshire, College Lane, Hatfield, AL10 9AB, UK
\and National Research Council of Canada, Herzberg Astronomy and Astrophysics, 5071 West Saanich Road, Victoria, BC V9E 2E7, Canada
\and Department of Astronomy and Space Science, Sejong University, Seoul 05006, Korea
\and Department of Physics, The Chinese University of Hong Kong, Shatin, N.T., Hong Kong S.A.R., China
\and Aix-Marseille Univ., CNRS, CNES, LAM, Marseille, France
\and University of Calgary, 2500 University Drive NW, Calgary, Alberta, T2N 1N4, Canada
\and Universit\`e de Strasbourg, CNRS, Observatoire astronomique de Strasbourg (ObAS), UMR 7550, F-67000 Strasbourg, France
\and UK Astronomy Technology Centre, Royal Observatory, Blackford Hill, Edinburgh, EH9 3HJ, UK
\and Yunnan Observatories, Chinese Academy of Sciences, Kunming 650011, People's Republic of China
\and State Key Laboratory of Radio Astronomy and Technology, National Astronomical Observatories, Chinese Academy of Sciences, Beijing 100101, China
\and Department of Astronomy, Yonsei University, Seoul, 03722, Republic of Korea
\and Center for Galaxy Evolution Research, Yonsei University, Seoul, 03722, Republic of Korea
\and NSF NOIRLab, 950 N. Cherry Avenue, Tucson, AZ 85719, USA
\and Department of Astronomy, Peking University, Beijing 100871, China
\and Kavli Institute for Astronomy and Astrophysics, Peking University, Beijing 100871, People's Republic of China
\and Universit\`a di Milano-Bicocca, Piazza della Scienza 3, 20100 Milano, Italy
\and AIM, CEA, CNRS, Universit\`e Paris-Saclay, Universit\`e Paris-Diderot, Sorbonne Paris-Cit\`e, Observatoire de Paris, PSL University, 91191 Gif-sur-Yvette Cedex, France
}

\date{June 2025}

\abstract
{Dwarf-dwarf galaxy mergers are among the least explored aspects of dwarf galaxy pre-processing as they fall into clusters.}
{We present the first case study of a coalesced late-type dwarf major merger (VCC 479; stellar mass $\sim\,8\,\times\,10^7\,\rm M_\odot$) that has undergone significant environmental influence, with the aim of exploring dwarf galaxy evolution under the combined effects of galaxy interactions and environmental processes, and understanding its relevance to the diversity of dwarf galaxies in cluster environments.}
{Our analysis is based on multifrequency observations, including VLA and FAST HI emission line mapping from the Atomic gas in Virgo Interacting Dwarf galaxies (AVID) survey, and existing ultraviolet and optical images. We also performed idealized hydrodynamical simulations of dwarf-dwarf mergers to help us interpret the observations.}
{We identify symmetric stellar shell structures in VCC 479, indicative of a coalesced major merger of dwarf galaxies. The galaxy features a central starburst, initiated $\sim$ 600 Myr ago, embedded within an exponential disk quenched $\sim$ 1 Gyr ago. The starburst contributes only 2.9$\pm$0.5\% of the total stellar mass, and VCC 479's global star formation rate is 0.3 dex lower than typical dwarfs of a similar mass. The galaxy is highly HI-deficient, with most HI gas concentrated within the central 1 kpc and little extended HI envelope. The misalignment of the HI velocity field with the stellar body is best explained by merger-triggered gas inflow, as is seen in our simulations.
}
{Our analysis is consistent with a scenario in which the majority of HI gas of the progenitor galaxies was removed by the cluster environment prior to the final coalescence. The merger concentrates the remaining gas toward the galaxy center, triggering a central starburst. The combined effect of environment stripping and galaxy merger has transformed VCC 479 into a blue-core dwarf undergoing a morphological transition from a late-type to an early-type galaxy.}

\keywords{Dwarf galaxies; Galaxy merger; HI gas; Starburst galaxies; Galaxy evolution}

\maketitle

\section{Introduction}
Dwarf galaxies represent the most numerous subset of galaxies in the Universe \citep{Binggeli1988}. Within the framework of the standard $\Lambda$ cold dark matter ($\Lambda$CDM) cosmological model, galaxies form hierarchically, which occurs through hierarchical mergers or accretion of smaller galaxies \citep[e.g.,][]{Annibali2019}. Galaxy mergers play a pivotal role in the formation and evolution of galaxies. Observations and numerical simulations of mergers between massive galaxies consistently demonstrate their significant impact on galaxy morphology evolution \citep[e.g.,][]{1991ApJ...370L..65B,1994ApJ...431L...9M,Lotz2008,Bournaud2011}, the enhancement of star formation \citep[e.g.,][]{1996ApJ...464..641M,Zhang2010,Teyssier2010,Cibinel2019}, and the triggering of starbursts and active galactic nuclei through gas inflows \citep[e.g.,][]{Mihos1995,Ellison2011,Patton2011,Torrey2012,Weston2017}, particularly in gas-rich major mergers.

Dwarf galaxies are building blocks of massive galaxies and are expected to be more common than their massive counterparts. The cosmological zoom-in simulations by \citet{Deason2014} concluded that 10\% - 20\% of Local Group satellite galaxies have experienced a major merger event since z = 1. Some studies suggest that dwarf mergers play a significant role in triggering star formation in the nearby Universe \citep{Kravtsov2013}. However, galaxy mergers at the low-mass end may be different due to the shallow gravitational potential wells. Observationally, fewer than 2\% of dwarf galaxies in the field are quiescent. This has led to the suggestion that mergers between dwarf galaxies do not quench the galaxies, in contrast to the typical outcome of major mergers involving more massive galaxies \citep{Bekki1998,Hopkins2008,Ellison2018}.

Unfortunately, studies of dwarf-dwarf mergers remain scarce in the literature, and the general impact of such mergers on their star formation, structure, and gas properties is still not understood. This is partly due to the fact that these faint systems are more challenging to identify compared to their massive counterparts. Only with the advent of low-surface-brightness imaging techniques have we been able to better detect such systems via faint, merger-induced features such as shells and tails \citep[e.g.,][]{2014PASP..126...55A}. Consequently, previous studies on dwarf mergers have primarily focused on well-separated pairs. \citet{Stierwalt2015} systematically studied interacting dwarf-dwarf pairs for the first time, focusing on star formation enhancement triggered by dwarf galaxy interactions. They found that the star formation rate (SFR) of dwarf pairs is typically enhanced by a factor of $\sim$ 2, which is similar to massive galaxy pairs \citep{Ellison2013,Silva2018}. More recently, \citet{Huang2024} quantified the interaction stages of dwarf pairs and reported that the SFR is suppressed during early stages of interactions, just after the halo encounter. Regarding the gas properties of dwarf pairs, studies indicate that they are gas-rich in isolated environments \citep{Stierwalt2015}. Additionally, \citet{Pearson2016} demonstrated that dwarf pairs exhibit more extended atomic gas distributions than isolated ones. Similar findings were also reported for massive galaxy pairs \citep[e.g.,][]{Wang2025}. \citet{Kim2023} investigated the impact of the cluster environment on the HI kinematics of dwarf pairs, finding that it reduces the fraction of kinematically narrow HI gas, increases gas stability, and enhances HI morphological disturbances. In addition to the aforementioned sample-based studies, case studies of individual dwarf pairs have been conducted \citep[e.g.,][]{Annibali2016,Paudel2017a,Privon2017,Makarova2018,Paudel2018a,Johnston2019,Elson2022,Gao2022,Gao2023,Liu2023,Zhang2024}.

For the dwarf-dwarf merger remnants, studies of their star formation activities and gaseous and stellar distributions are rare due to the difficulty of identifying them. \citet{Amorisco2014} provided the first observational evidence of a dwarf-dwarf merger remnant, revealing a stellar stream in Andromeda II. Based on the Next Generation Virgo Cluster Survey (NGVS; \citealt{Ferrarese2012}) optical image, \citet{Paudel2017} identified three early-type dwarfs with shells in Virgo, which are remnants of gas-free major mergers. \citet{Zhang2020a,Zhang2020b} reported the first confirmed coalesced gas-rich major merger VCC848, and found that star formation during the merger was enhanced by a factor of 10 or so and the merger remnant has similar stellar structures to the ones of ordinary dwarfs.
Beyond these confirmed cases, other studies have compiled samples of potential merger remnants. \citet{Paudel2018a} compiled a catalog of 177 relatively low-mass merger candidates based on SDSS-III and the Legacy Survey. Among these, 30 exhibit shell features, and 49 display tidal tails, classifying them as post-merger systems. Their analysis revealed that the sample is predominantly composed of star-forming galaxies, with several identified as blue compact dwarfs (BCDs) \citep{Chhatkuli2023}. Similarly, \citet{Kado-Fong2020} constructed a dwarf merger sample by detecting tidal features in optical images. They found evidence that dwarf galaxies exhibiting starburst activity are more likely to display tidal debris, concluding that merger-driven star formation persists after coalescence. Their results suggest that mergers between dwarf galaxies might trigger starbursts and form BCDs. In simulations, gas-rich dwarf-dwarf mergers have been proposed as a mechanism for forming BCDs by triggering gas inflows and central starbursts \citep{Bekki2008}.
Despite these findings, it remains unclear what the general consequence of dwarf-dwarf mergers is, especially the gas-bearing ones, for the gas and stellar distribution of dwarf galaxies, as well as what outcomes arise from the combined effects of the environment and the merger process.

In this work, we investigate the stellar and HI gas properties of the dwarf post-merger VCC 479, one of the galaxies within the AVID (Atomic gas in Virgo Interacting Dwarfs) sample, located on the outskirts of the Virgo cluster. We selected this galaxy because of its striking symmetric shell structure, which strongly suggests a past major merger event (Section \ref{sec:shell}). Studying this system provides a valuable opportunity to investigate how major mergers shape the evolution of dwarf galaxies in a cluster environment. Specifically, we aim to examine the effects of such mergers on star formation, gas content, structural properties, and overall evolutionary processes. We utilize multifrequency observational data, including deep optical imaging data from NGVS, interferometric and single-dish HI gas observations, as well as other auxiliary datasets. Some pertinent properties of VCC 479 are summarized in Table \ref{tab:para}. The structure of this paper is as follows: In Section \ref{sec:data}, we describe the data utilized in this study. In Section \ref{sec:result}, we present the analysis and results from UV-optical data. In Section \ref{sec:HI}, we present the results of HI observations. In Section \ref{sec:simulation}, we introduce numerical simulation to interpret our observations. We discuss the implications from our results for dwarf-dwarf mergers in Section \ref{sec:discussion}. Finally, a summary is provided in Section \ref{sec:sum}. Throughout this paper, we adopt a distance of 16.5 Mpc for VCC 479 (the distance of Virgo given by \citet{Mei2007}).

\begin{table}
\caption{Properties of VCC 479}
\label{tab:para}
\centering
\begin{tabular}{lc}
\hline\hline
Property & Value \\
\hline
$^a$RA & 12h21m29.00s \\
$^a$Dec & 8d9m3.80s \\
$^a$PA & 61° \\
$^a$Ellipticity & 0.51 \\
$^b$Distance & 16.5 Mpc \\
Heliocentric radial velocity & 2531 km s$^{-1}$ \\
Angular distance from M87 & 4.8° \\
Absolute B magnitude & -14.55 \\
Absolute \textit{g} magnitude & -14.85 \\
(\textit{g}$-$\textit{i}) color & 0.56 mag \\
12 + log(O/H) & 8.22 \\
R$_{e,g}$ & 17.0 arcsec \\
$\mu_{0,g}$ & 21.5 mag arcsec$^{-2}$ \\
Stellar mass & $8.0\times10^7$ M$_\odot$ \\
HI mass & $1.4\times10^7$ M$_\odot$ \\
H$\alpha$ luminosity & $3.9\times10^{38}$ erg s$^{-1}$ \\
$^c$SFR$_{\rm FUV,corr}$ & 0.0039 M$_\odot$ yr$^{-1}$ \\
$^c$SFR$_{\rm H\alpha,corr}$ & 0.0026 M$_\odot$ yr$^{-1}$ \\
\hline
\end{tabular}
\tablefoot{
$^a$ Center, PA, ellipticity are all derived based on our \texttt{ellipse} fitting in Section \ref{sec:SB}.
$^b$ Taken as the distance of Virgo in \citet{Mei2007}.
$^c$ SFR derived from the extinction-corrected FUV and extinction-corrected H$\alpha$ luminosity.
}
\end{table}

\section{Observations and data reduction} \label{sec:data}

\subsection{VLA multi-configuration HI emission line observations}
VCC 479 was observed in B (Dec 2021), C (June 2021), and D (April 2021) array configurations of the NRAO\footnote{The National Radio Astronomy Observatory is a facility of the National Science Foundation operated under cooperative agreement by Associated Universities, Inc.} Karl G. Jansky Very Large Array \citep[VLA;][]{Perley2011} radio interferometer as part of the AVID project (program ID: 21A-231; PI: Hong-Xin Zhang). HI spectral line observations of the VCC 479 field were performed with a 16 MHz bandwidth and a 1.65 km $\rm s^{-1}$ channel separation. The total on-source time of observations is 4.6 hr, 1.94 hr and 1.03 hr respectively in B, C and D configurations. The multiconfiguration datasets were calibrated, combined, and imaged using the CASA software package \citep{McMullin2007}.
Details of the observations and data reductions of AVID galaxies are presented in Zhang et al. (in prep). The image cubes are deconvolved down to 1$\times$RMS (root mean square) noise level with the \textsc{tclean} task in CASA. We use HI image cubes produced through two different $uv$-data weighting schemes: one is the natural weighting (hereafter natural-weighted cube) and the other one is a tapered natural weighting by applying a Gaussian taper to the natural weights with a Gaussian half-width at half-maximum of 12k$\lambda$ in the $uv$-plane (hereafter tapered natural-weighted). The natural-weighted image cube has a beam size of 12.7 $\times$ 9.4 arcsec (pixel size = 2.0 arcsec) and an RMS noise of 0.51 mJy beam$^{-1}$, corresponding to a minimum detectable HI column density of $\sim5.7\,\times\,10^{19}\,\rm cm^{-2}$ (3 $\sigma$ over 10 km s$^{-1}$), while  the tapered image cube has a beam size of 21.7 $\times$ 17.3 arcsec (pixel size = 4.0 arcsec) and an RMS noise of 0.60 mJy beam$^{-1}$, corresponding to a minimum detectable HI column density of $\sim2.1\,\times\,10^{19}\,\rm cm^{-2}$ (3 $\sigma$ over 10 km s$^{-1}$). The HI cubes are Hanning smoothed to have an effective spectral resolution of 3.3 km $\rm s^{-1}$. In this work, we utilize the natural-weighted cubes to explore the HI spatial distribution and velocity field, and utilize the tapered cube to derive the spatially integrated HI flux and compare it with single-dish observations to assess potential flux loss in the VLA data. The HI emission was detected by running the Source Finding Application (SoFiA-2; \citealt{Serra2015,Westmeier2021}) on the HI cubes. We adopted the SoFiA smooth-and-clip algorithm (S+C) to search for genuine HI emission in the image cubes, with spatial smoothing kernels of 0 and $\sim$ 1.5 $\times$ the average beam size, spectral smoothing kernels of 0, 3 and 11 channels, and a detection threshold of 4 $\times$ RMS. Based on the S+C detection, moment 0, 1, and 2 maps were generated, representing integrated HI distribution, intensity-weighted velocity field, and intensity-weighted velocity dispersion distribution.

\subsection{FAST HI mapping}
VCC 479 was observed with the Multibeam on-the-fly (OTF) mode of the Five-hundred-meter Aperture Spherical Telescope (FAST) to map the HI distribution and reveal potential diffuse HI features that may fall below the detection limit of the VLA observations. The OTF mapping covers a $\sim$ 30 $\times$ 30 arcmin sky area centered on VCC 479. Each observing session consists of a pair of horizontal and vertical scans of the sky area. The total observing time is $\sim$ 1900 seconds for a single pair of vertical and horizontal scanning observation. Observations were conducted on four separate days: August 11, 12, 16, and September 20, 2022. Observations from September 20 were severely affected by standing waves and were excluded from the analysis. The accumulated on-source time per sky position is about 220 seconds. The observing strategy follows the approach described in \citet{Wang2023} and \citet{Yang2025}. We chose the Spec(W+N) spectrometer, which offers 65,536 channels and covers a bandwidth of 500 MHz for each polarization and beam, with a velocity spacing of 1.61 km s$^{-1}$. The raw spectral data, collected using the 19-beam L-band receiver, were processed with the HiFAST pipeline\footnote{\url{https://hifast.readthedocs.io}} \citep{Jing2024}. This dedicated, modular, and self-contained package handles various steps in single-dish spectral line data reduction, including temperature calibration, baseline subtraction, standing wave removal, radio frequency interference (RFI) flagging, gridding, and smoothing. The beam FWHM of the image cube produced by HiFAST is 3.24 arcmin. The final image cube has an RMS noise level of approximately 1.24 mJy beam$^{-1}$ per channel, corresponding to a minimum detectable HI column density of $\sim4.4\,\times\,10^{17}\,\rm cm^{-2}$ (3 $\sigma$ over 10 km s$^{-1}$). As with the VLA data, genuine HI emission detected by FAST was also searched by using the SoFiA software with two spatial kernels (0 and 3 arcmin) and three spectral kernels (0, 3 and 11 channels), with a flux threshold of four times the RMS noise.

\subsection{Other multiwavelength data}

\subsubsection{Broadband optical imaging}
Broadband optical images of VCC 479 in the \textit{u}, \textit{g}, \textit{i}, \textit{z} bands are available from Next Generation Virgo Survey (NGVS), which observed the entire Virgo cluster with the MegaCam on the Canada-France-Hawaii Telescope (CFHT) \citep{Ferrarese2012}. The images were reduced with the Elixir LSB pipeline and a subsequent global background subtraction and artificial skepticism combine \citep{Ferrarese2012}. The point-spread function (PSF) full width at half maximum (FWHM) is 0.5$-$0.6 arcsec in the \textit{i} band, and the 2$\sigma$ surface brightness limit is $\mu_g\,\sim\,$ 29 mag arcsec$^{-2}$, with a pixel scales of 0.186 arcsec. These data enable the detection of extended, low-surface-brightness features such as shells and tidal tails. A surface brightness analysis of the deep optical images of VCC 479 is presented in Section \ref{sec:SB}.

\subsubsection{VESTIGE narrow-band H$\alpha$ imaging}
Narrow-band H$\alpha$ images were obtained from the Virgo Environmental Survey Tracing Ionised Gas Emission (VESTIGE), a deep survey of the Virgo cluster also with MegaCam on the CFHT \citep{Boselli2018a}. The narrow-band H$\alpha$
filter has a center wavelength of $\lambda_c$ = 6591 \text{\AA} and a bandwidth of $\Delta\lambda$ = 106 \text{\AA}. Broad-band r-filter observations are also conducted and are used to subtract the stellar continuum. The data reduction process is similar to that applied to the NGVS images. The resulting sensitivity is $f(\rm H\alpha)\,\sim\,4\times10^{-17}\rm erg\,\rm s^{-1}\,cm^{-2} (5\sigma)$ for point sources and $\Sigma(\rm H\alpha)\,\sim\,2\,\times\,10^{-18}\,\rm erg\,\rm s^{-1}\rm cm^{-2}\,\rm arcsec^{-2}$ (1$\sigma$ after smoothing the data to $\sim$3 arcsec resolution) for extended sources. The typical seeing FWHM is 0.76 arcsec.
\subsubsection{GALEX UV imaging}
FUV and NUV images were obtained from the GALEX (Galaxy Evolution Explorer) Ultraviolet Virgo Cluster Survey (GUViCS) \citep{Boselli2011}. The angular resolution of the NUV band image is approximately $\sim$ 5 arcsec. UV data are sensitive to the distribution of young stars with an age $\lesssim100$ Myr \citep{Kennicutt1998}, and will be used to investigate the recent star formation in VCC 479.

\subsubsection{SDSS single-fibre spectra}
An SDSS spectrum of the central starburst region of VCC 479 was obtained using a 3-arcsecond diameter fiber positioned at the focal plane of the 2.5-meter telescope at the Apache Point Observatory \citep{2009ApJS..182..543A}. The spectral coverage spans a wavelength range of 3800 to 9200 \text{\AA}, with an average spectral resolution of $\lambda/\Delta\lambda\,\sim\,$1800.

\section{Analysis and results from UV-optical observations} \label{sec:result}
\subsection{Optical appearance of VCC 479}\label{sec:shell}
\begin{figure}
    \centering
    \includegraphics[width=0.99\linewidth]{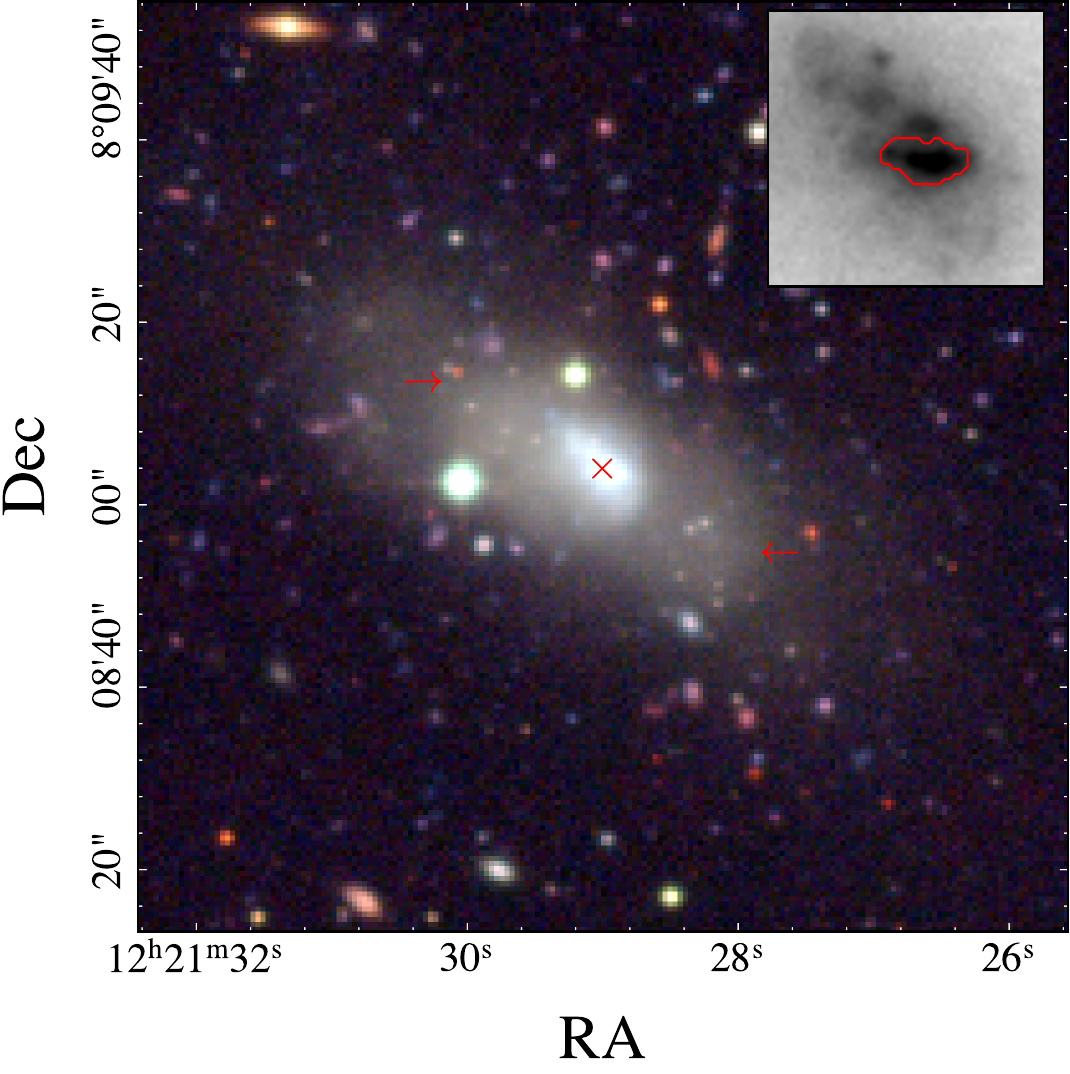}
    \caption{Color composite image made from NGVS \textit{u}-, \textit{g}-, and \textit{i}-band data. The red cross represents the galaxy center. We indicate the location of the stellar shells with two red arrows. In the top right corner, we present a zoomed-in \textit{i}-band image of the central 10 arcsec region. The brightest star clusters detected by \texttt{photutils} are delineated with a red contour.}
    \label{fig:shell}
\end{figure}

\begin{figure*}
    \centering
    \includegraphics[width=0.32\linewidth]{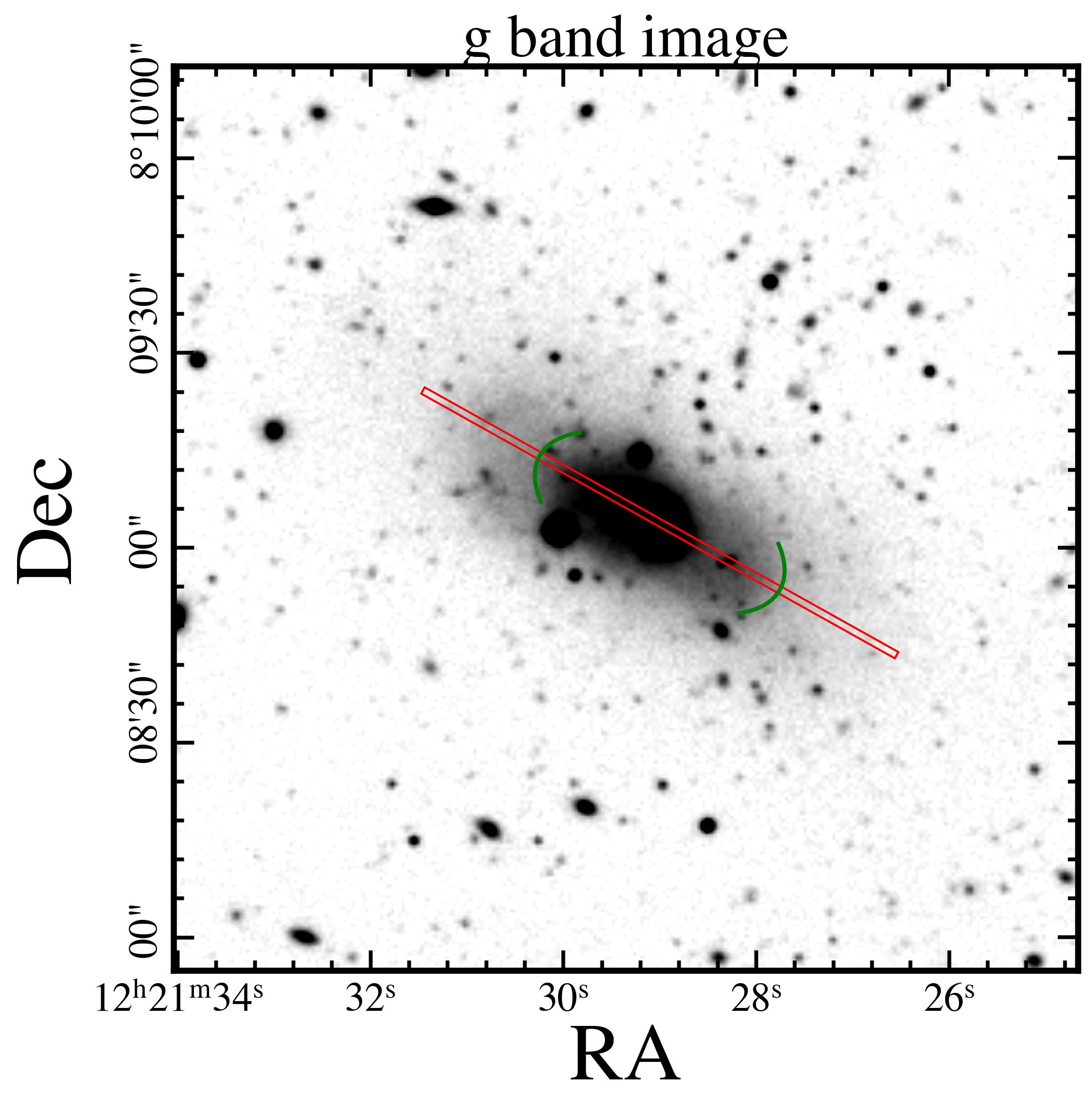}
    \includegraphics[width=0.33\linewidth]{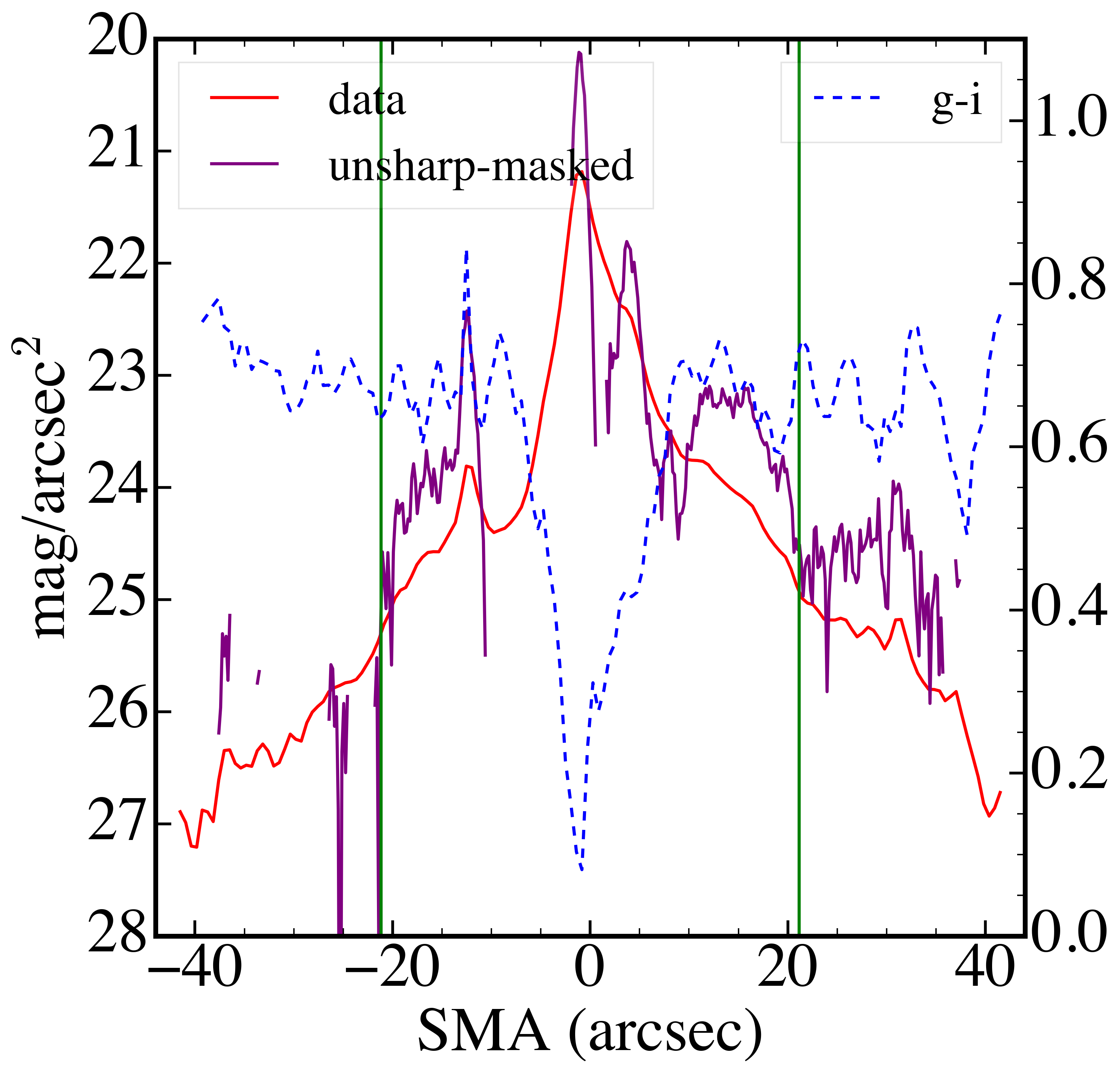}
    \includegraphics[width=0.32\linewidth]{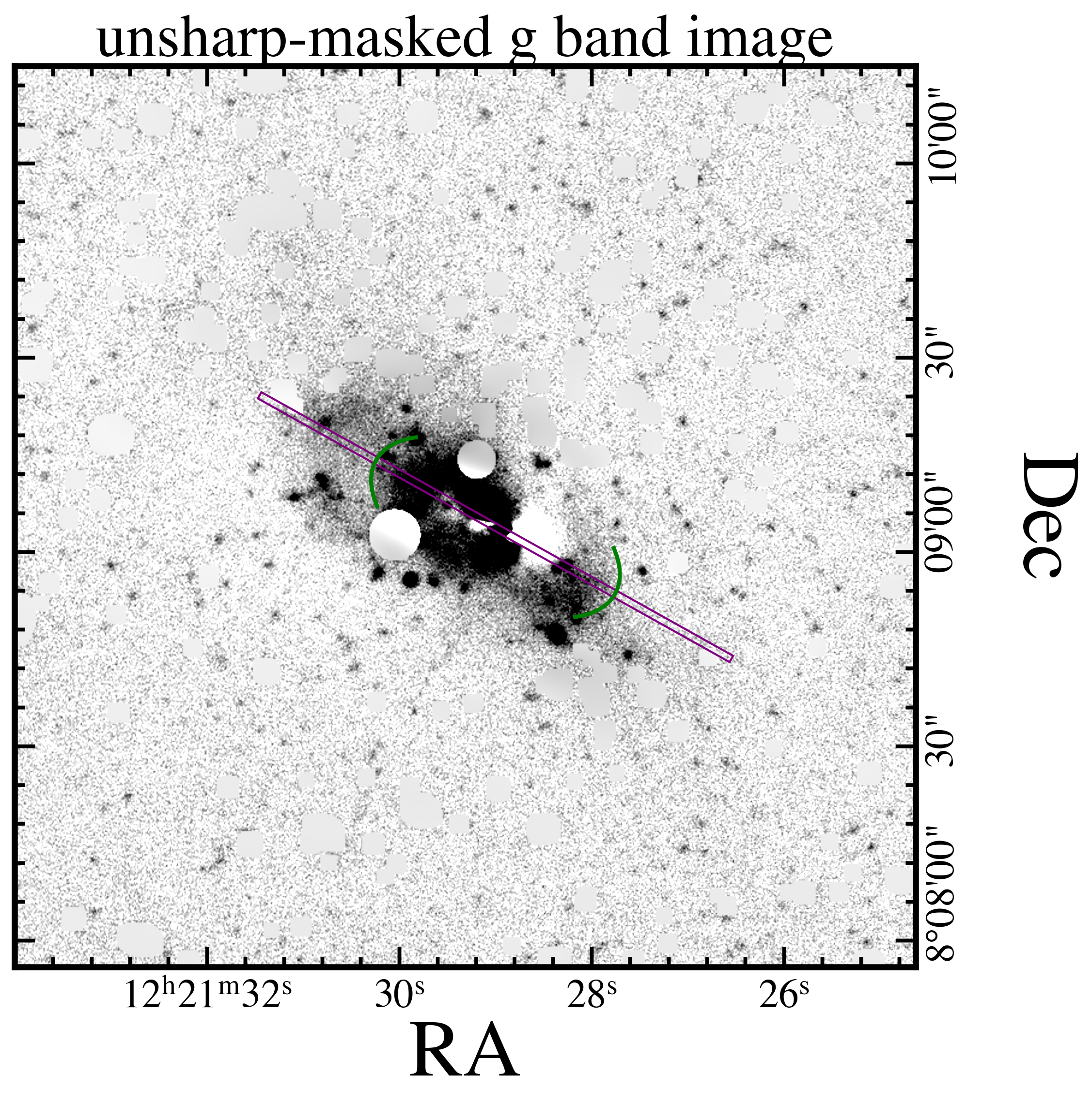}
    \caption{Left panel: NGVS \textit{g}-band image in grayscale. The green ellipse arcs mark the shells. Middle panel: Slit profile of original \textit{g}-band image (solid red curve) and unsharp-masked image (solid purple curve). The \textit{g$-$i} color slit-profile measured from the smoothed $g$ and $i$ images by ADAPTSMOOTH \citep{Zibetti2009} is plotted with a dashed blue line. The green vertical lines mark the radii of the stellar shells. Right panel: Unsharp-masked \textit{g}-band image. The red (left panel) and purple (right panel) rectangular regions indicate the area used to extract the slit profiles shown in middle panel.}
    \label{fig:slit-profile}
\end{figure*}

In Figure \ref{fig:shell}, prominent symmetric shell structures is visible along the galaxy's major axis in the \textit{u}, \textit{g}, and \textit{i} color composite image. Typically, a break in the surface brightness map that forms an arc-like shape is referred to as a shell. To better illustrate this feature, we extracted slit surface brightness profiles from the NGVS \textit{g} band image (left panel of Figure \ref{fig:slit-profile}) and its unsharp-masked image (right panel of Figure \ref{fig:slit-profile}) along the major axis. The middle panel of Figure \ref{fig:slit-profile} exhibits the slit profile of VCC 479. Within the shell radius (marked by the green arc-like line), the surface brightness declines rapidly, whereas outside the shell, the decrease is more gradual. This contrast is more pronounced in the slit surface brightness profile derived from the unsharp-masked image. Additionally, the slit surface brightness profile reveals a symmetric shell structure with similar \textit{g}$-$\textit{i} color, suggesting that VCC 479 is the remnant of a recent major merger \citep{Paudel2017}.

\subsection{Local environment of VCC 479} \label{sec:environ}
\begin{figure}
    \centering
    \includegraphics[width=\columnwidth]{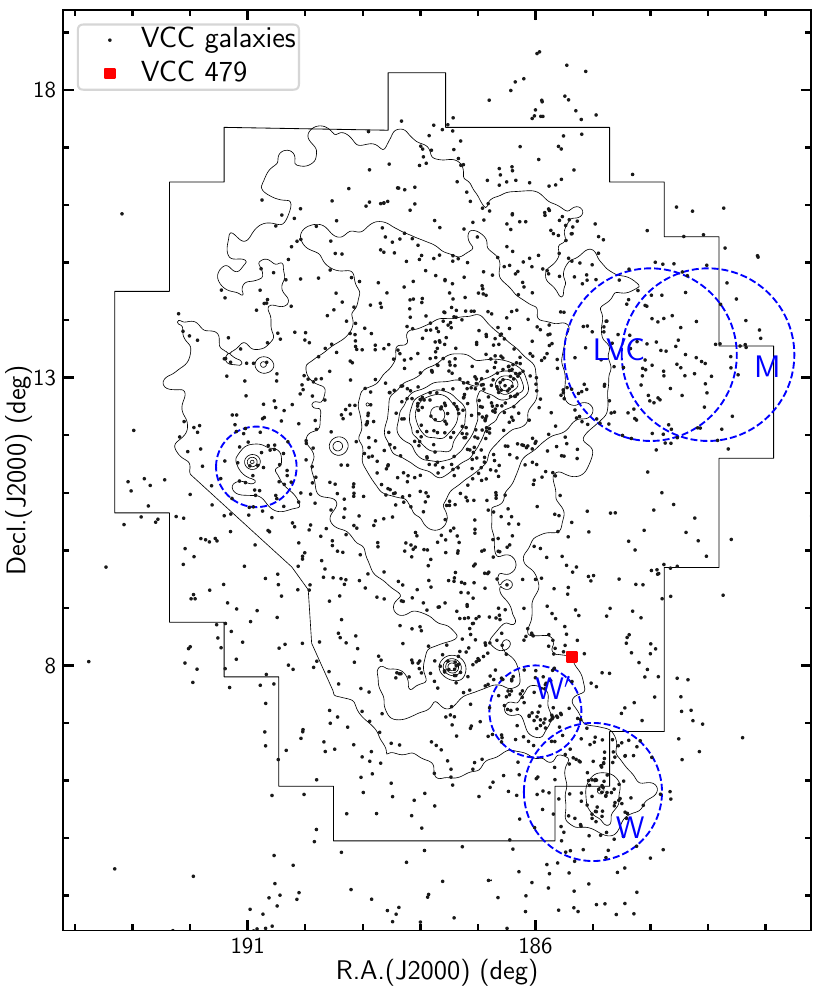}
    \caption{Spatial location of VCC 479 in Virgo cluster. The area in black shows the footprint of the NGVS. We denote the substructure identified by \citet{Boselli2014} as dashed blue cycles. The black contour indicate the X-ray diffuse emission of the cluster, from \citet{1994Natur.368..828B}.}
    \label{fig:AVID-virgo}
\end{figure}

\begin{figure*}
    \centering
    \includegraphics[width=0.40\linewidth]{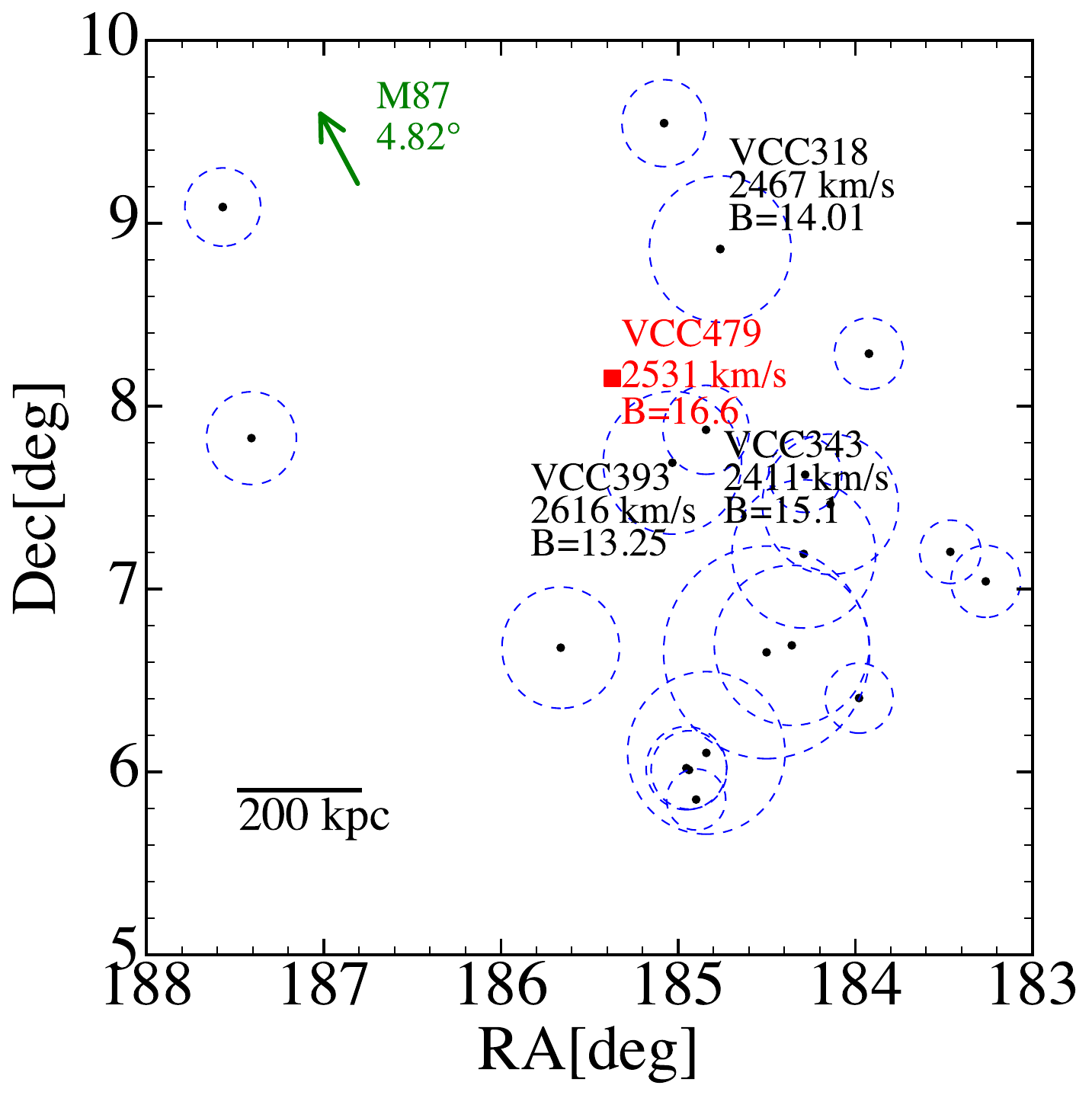}
    \includegraphics[width=0.59\linewidth]{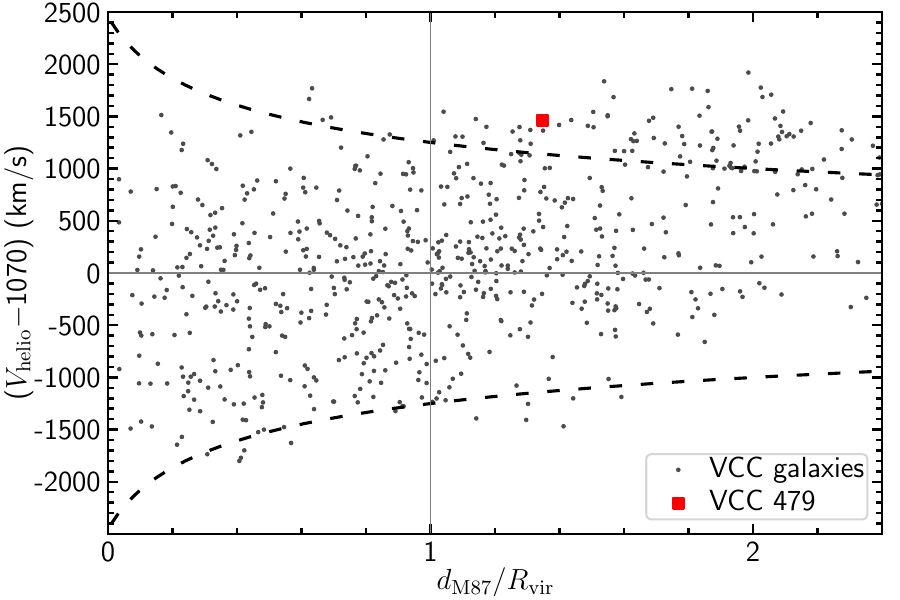}
    \caption{Left panel: VCC galaxies around VCC 479 within 700 kpc and $\pm$700 km s$^{-1}$. The dashed blue circles mark the virial radii of individual confirmed VCC galaxies. The virial radius is estimated as 67 times the r-band half-light radius, as provided in the EVCC catalog \citep{Kim2014}, following the method described in \citet{Kravtsov2013}. The magnitude and velocities of the three galaxies closest to VCC 479 (VCC 318, VCC 393, and VCC 343) within 300 kpc and 200 km s$^{-1}$ are indicated. The arrows denote the direction and angular distance of M87. Right panel: Projected phase-space diagram of redshift-confirmed Virgo galaxies. The two dashed black curves indicate the variation in the escape velocities of Virgo with the angular distance (normalized by R$_{200}$) from the Virgo cD galaxy M87. VCC 479 is located on the outskirts of Virgo.}
    \label{fig:AVID}
\end{figure*}

We show the location of VCC 479 along with other galaxies in the Virgo cluster in Figure \ref{fig:AVID-virgo}. VCC 479 is located (in projection) on the outskirts of the Virgo cluster, $\sim$ 4.8$\degree$ away from the M87. Based on the substructure identification of Virgo in
\citet{Boselli2014} (their Figure 1), we find that VCC 479 does not belong to any substructure but is still within the radius that separates field galaxies from cluster galaxies as defined by \citet{Boselli2014}. The left panel of Figure \ref{fig:AVID} shows the distribution of VCC galaxies within 700 kpc in projected radius and $\pm$ 700 km $\rm s^{-1}$ in radial velocity of VCC 479. The spatial and velocity ranges are chosen to be twice the projected virial radius and velocity dispersion of a typical group with a central galaxy of stellar mass of 10$^{10} M_\odot$ \citep{Makarov2011}. The right panel of Figure \ref{fig:AVID} displays the phase-space diagram, showing the line-of-sight velocity difference between individual galaxies and the average velocity (1070 km $\rm s^{-1}$ by \citet{Kashibadze2020}) of the Virgo cluster, plotted against the galaxies’ projected distance from the cluster center. The distance has been normalized by the virial radius of Virgo cluster (0.998 Mpc by \citet{2017MNRAS.469.1476S}). VCC 479 lies above the cluster escape velocity curve, suggesting it is most likely recent infalling (Figure \ref{fig:AVID}).
It is clear that VCC 479 is free from significant gravitational influence by neighboring galaxies, given its current location and radial velocity. We note that, in the group catalog based on SDSS data by \citet{Tempel2017}, VCC 479 is classified as an isolated galaxy. Nevertheless, the possibility that VCC 479 had close interactions in the past with galaxies such as VCC 393 cannot be ruled out.

\subsection{Isophotal analysis and UV-optical photometry} \label{sec:SB}
\begin{figure}
    \centering
    \includegraphics[width=\columnwidth]{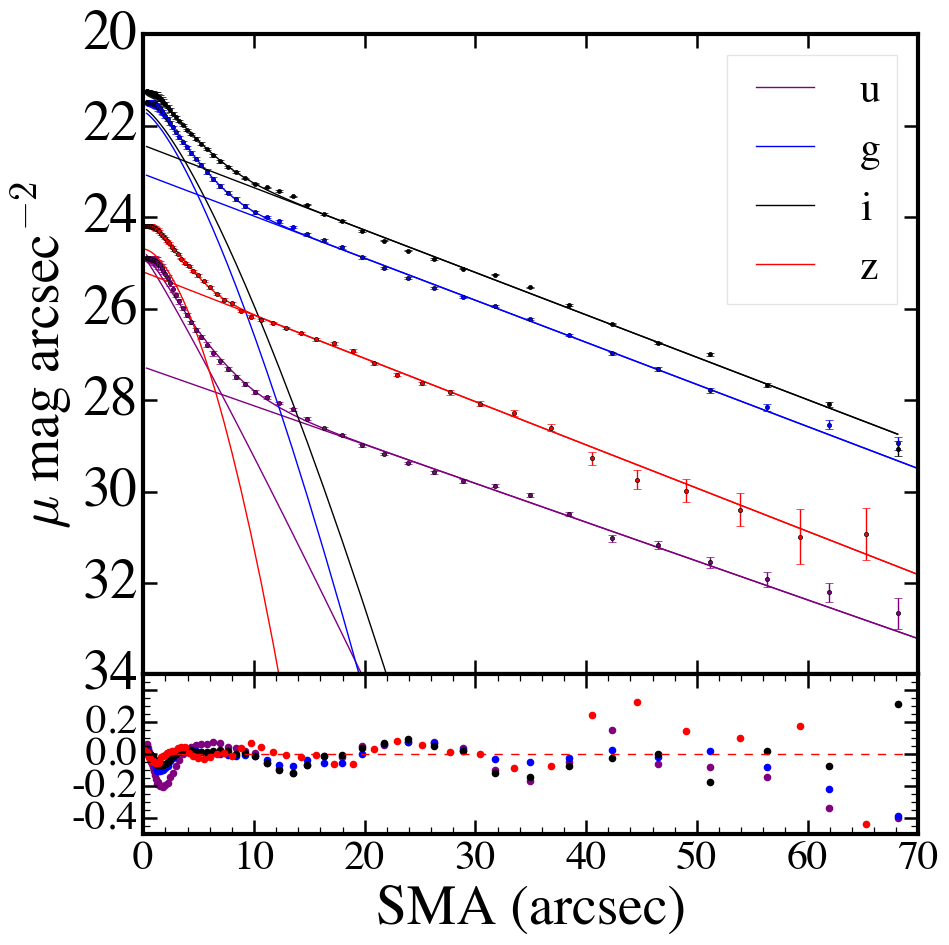}
    \caption{Surface brightness profiles of VCC 479 in the NGVS \textit{u}, \textit{g}, \textit{i}, and \textit{z} bands. The dots with error bars denote the average surface brightness profile from the \texttt{ellipse} fitting. Solid curves show the best-fit decomposition of the surface brightness profiles into an outer exponential component and a central light excess modeled with a Sersic profile. For clarity, the surface brightness profiles in the \textit{u} and \textit{z} bands are arbitrarily shifted downward by 3 mag arcsec$^{-2}$. The residuals, obtained by subtracting the best-fit profiles, are shown in the bottom panel.}
    \label{fig:SB}
\end{figure}

We performed surface photometry on the \textit{u-}, \textit{g-}, \textit{i-}, and \textit{z}-band images using the \texttt{ellipse} task from the \texttt{photutils} Python package. Foreground and background sources in the images were masked prior to performing photometry. The isophotal fitting was conducted iteratively. Specifically, we first determined the average geometric parameters, including the galaxy center, ellipticity (0.51), and position angle (PA; 61\degree), by fitting the g-band isophotal ellipses at large galactocentric radii that are beyond the central starburst area. Then, by fixing the geometric parameters, we extracted the surface brightness profile of the galaxy in the \textit{u}, \textit{g}, \textit{i}, and \textit{z} bands. The resulting profiles are shown in Figure \ref{fig:SB}. Next, we modeled the surface brightness profiles with a superposition of a central Sersi\'c function and an outer exponential function. In all bands, the two-component decomposition gives a decent fitting to the radial profiles, as shown in Figure \ref{fig:SB}. We note that the surface brightness profiles in the four bands show systematic positive residuals (lower panel of Figure \ref{fig:SB}) in 20-30 arcsec, which could be attributed to the presence of a shell. In the central 1 arcsec, a significant residual is present in the \textit{u} band, which is caused by the central starburst region. It is noteworthy that the overall exponential decline of surface brightness of VCC 479 is in line with the correlation between Sersic index (\textit{n}; \textit{n}=1 corresponds to an exponential profile) and absolute \textit{B}-band galaxy magnitude followed by ordinary galaxies \citep{Graham2003}. This dwarf merger remnant thus developed a disk-like structure, similar to our finding for the gas-rich merger remnant VCC 848 \citep{Zhang2020a}.

\subsection{Metallicity, dust reddening, and stellar populations from the SDSS single-fibre spectrum}\label{sec:SDSS}
\begin{figure*}
    \centering
    \includegraphics[width=\textwidth]{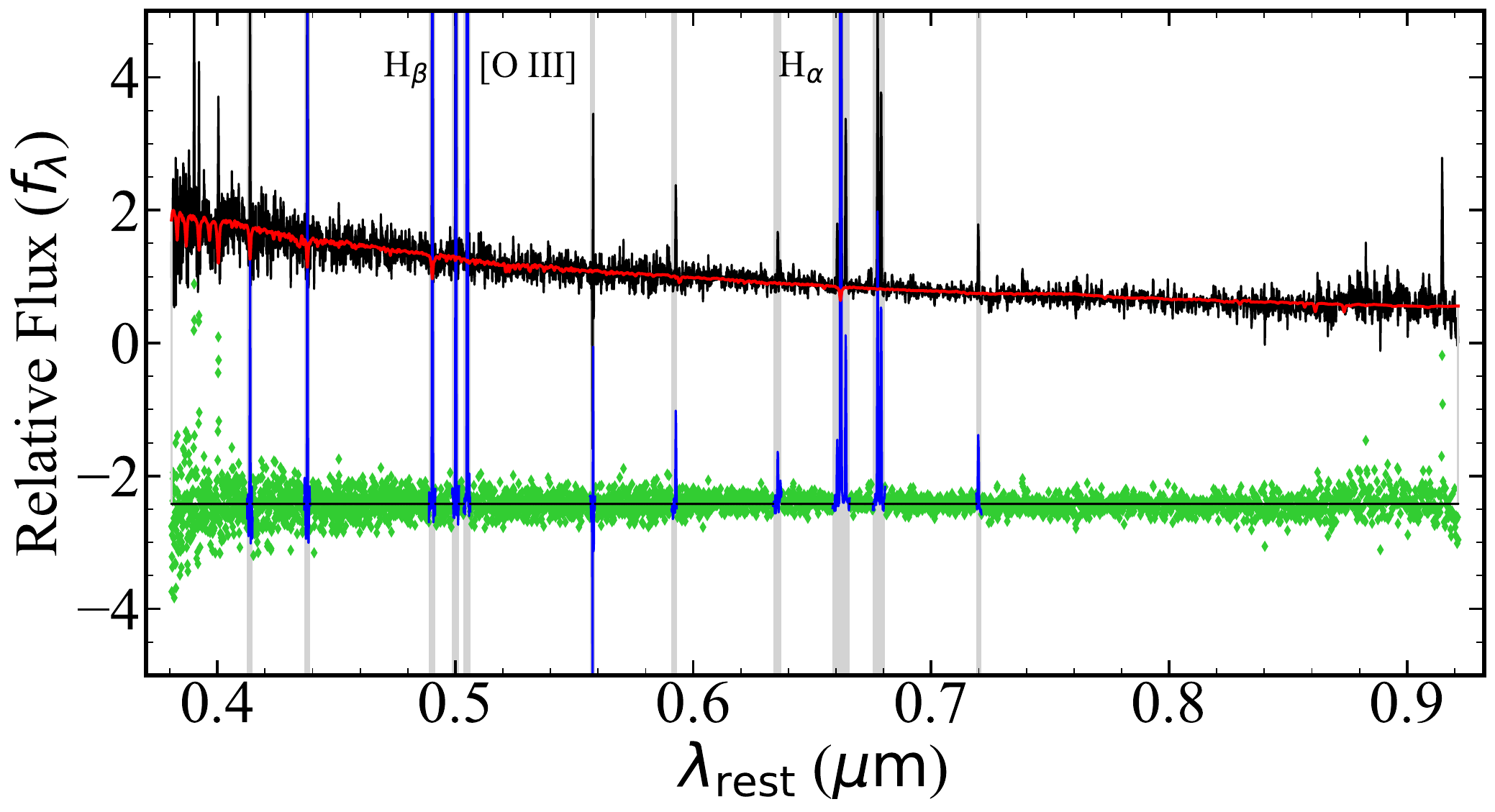}

    \caption{SDSS spectrum and its best-fitting result by pPXF. The black line represents the observed SDSS spectrum of VCC 479 The red line shows the best-fitting spectrum obtained by using pPXF fitted with \citet{bc03} libraries. The gray regions indicate the lines which have been masked before the fitting of the stellar continuum. The blue lines indicate emission lines. Those used to calculate metallicity and dust reddening are labeled. The green dots represent the residuals.}
    \label{fig:spec-SDSS}
\end{figure*}

VCC479 has an SDSS fiber spectrum observation in a central 3 arcsec aperture, probing the central starburst region. As expected, the spectrum exhibits strong emission lines, as shown in Figure \ref{fig:spec-SDSS}. These emission lines were used to calculate the internal attenuation and metallicity. The extinction E(B$-$V) = $A_V/4.05$, was estimated using the Balmer decrement, assuming an intrinsic emission-line flux ratio of $I_{\rm H\alpha}/I_{H\beta}$=2.86 \citep{Osterbrock2006}. To properly determine the emission-line strengths, we first removed the absorption line contributions from the underlying stellar populations by fitting the spectrum with penalized pixel-fitting (pPXF, \citep{Cappellari2004,Cappellari2017}). The best-fit spectrum was then subtracted from the observed spectrum. We obtained $I_{H\alpha}/I_{H\beta}$ = 3.1, which corresponds to E(B$-$V ) = 0.075 for internal gas extinction. This value was used to correct the VESTIGE H$\alpha$ narrow band flux and also as a prior information in our stellar population analysis in Section \ref{SED-global} and \ref{SED-resolve}. Note that we have applied Galactic extinction correction using the value E(B$-$V) = 0.021 provided in the GUViCS \citep{Voyer2014} catalog, assuming R$_{\rm V}$ = 3.1 and using the \citet{Fitzpatrick99} extinction law. This correction is also applied in the subsequent section for image photometry. The metallicity was estimated using the O3N2 method \citep{Pettini2004}. This gives 12+log(O/H) = 8.22, with a systematic error of 0.2 dex. The N2 method gives a consistent result (8.26). VCC 479 is consistent with the mass$-
$metallicity relation of dwarf galaxies from \cite{Jimmy2015} within 1$\sigma$ uncertainties. The stellar mass was obtained through SED (spectral energy distribution) fitting, as detailed in Section \ref{SED-resolve}.

We performed the full spectral fitting with pPXF to infer the star formation history (SFH). The emission lines were masked during the fitting. To maintain consistency with the photometric SED fitting (Section \ref{sec:SED-method}), we used the \citet{bc03} stellar population synthesis models as simple stellar population (SSP) templates, assuming a \citet{chabrier2003} IMF (initial mass function). The SSP templates span 43 ages from 0.001 Gyr to 15.85 Gyr. Given that the gas-phase metallicity derived from emission line is low (12+log(O/H) = 8.22), we excluded SSP templates with super-solar metallicity and only include four metallicities (i.e., Z = 0.0004, 0.004, 0.008, 0.02) \citep[e.g.,][]{Zhao2011}. The fitting was performed in two steps. In the first step, we set kinematics (velocity and velocity dispersion) as free parameters to obtain their initial values. A multiplicative Legendre polynomial (degree = 8, mdegree = 8) was applied to correct the continuum shape. The results from this step yield a velocity of 2593 km s$^{-1}$ and a velocity dispersion of 91 km s$^{-1}$. In the second step, the velocity and velocity dispersion were fixed to the values obtained in the first run.
Following the method suggested by \citet{Lee2024}, the pPXF reddening correction was used without any additional polynomial corrections (reddening keyword activated, degree = $-$1, mdegree = 0). This reddening correction is based on the \citet{Calzetti2000a} dust extinction law. The regularization parameter was set to 0.

\begin{figure}
    \centering
    \includegraphics[width=\columnwidth]{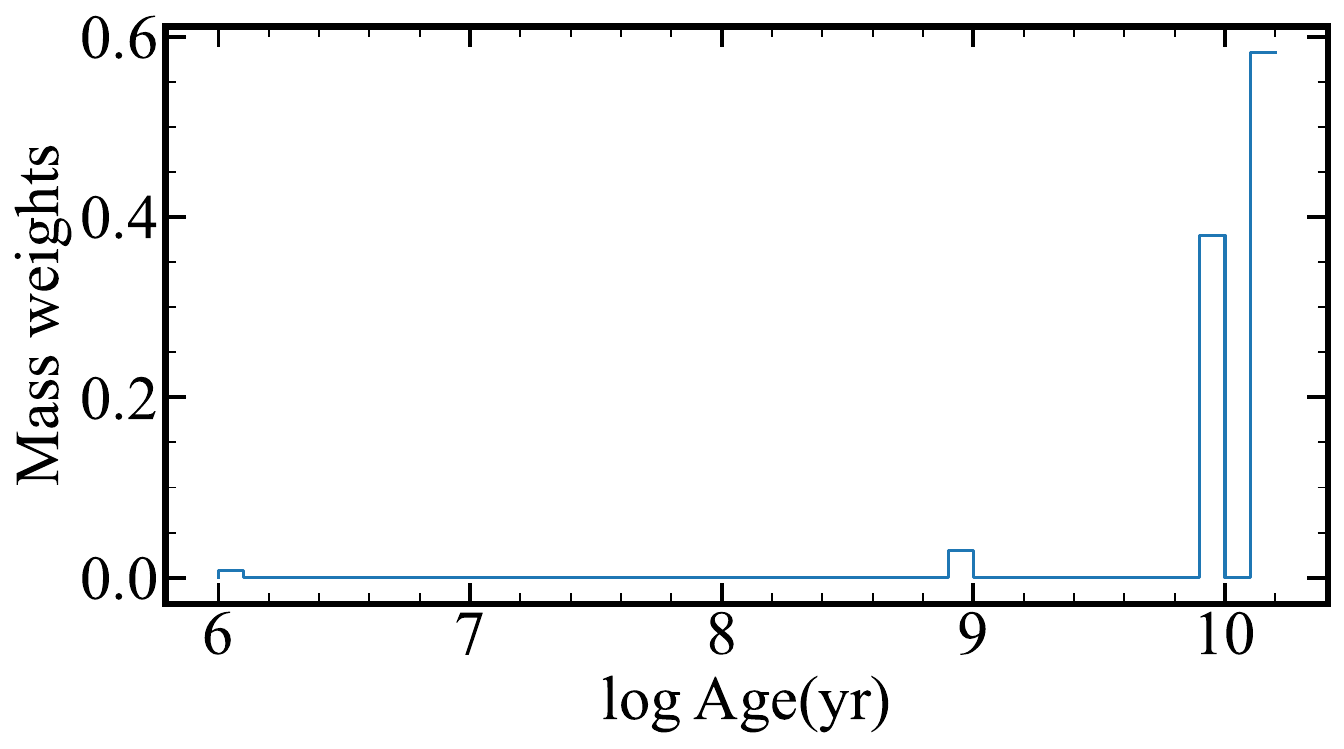}
    \caption{ Best-fit SFH by the output of the pPXF without regularization. It is given in terms of mass weights of SSPs with different ages.}
    \label{fig:spec-SFH}
\end{figure}

The best-fit spectrum is shown in Figure \ref{fig:spec-SDSS}, and the resulting SFH is shown in Figure \ref{fig:spec-SFH}. The stellar population is composed of old, intermediate, and young components. The old stellar population, resulting from the initial star formation, constitutes the vast majority of the stellar content. The combined mass fraction of the intermediate and young components, with ages younger than 1 Gyr, is 3.9\%. This result aligns well with \citet{Lisker2006a}, who fitted the SDSS spectra of a sample of blue-cored dwarf galaxies similar to VCC 479, and found a similar mass fraction for young stellar populations.

We verified the consistency between the spectral fitting results and the surface brightness profile decomposition. Specifically, the central excess of the $g$-band light above the inward exponential extrapolation accounts for approximately 75\% of the $g$-band light in the galaxy's central 3 arcsec (Figure \ref{fig:SB}). This is comparable to the $g$-band light contribution from stellar populations younger than 1 Gyr (65\%) derived from the above full-spectrum fitting.

\subsection{Star formation distribution}\label{SED-global}
\begin{figure*}
    \centering
    \includegraphics[width=0.31\linewidth]{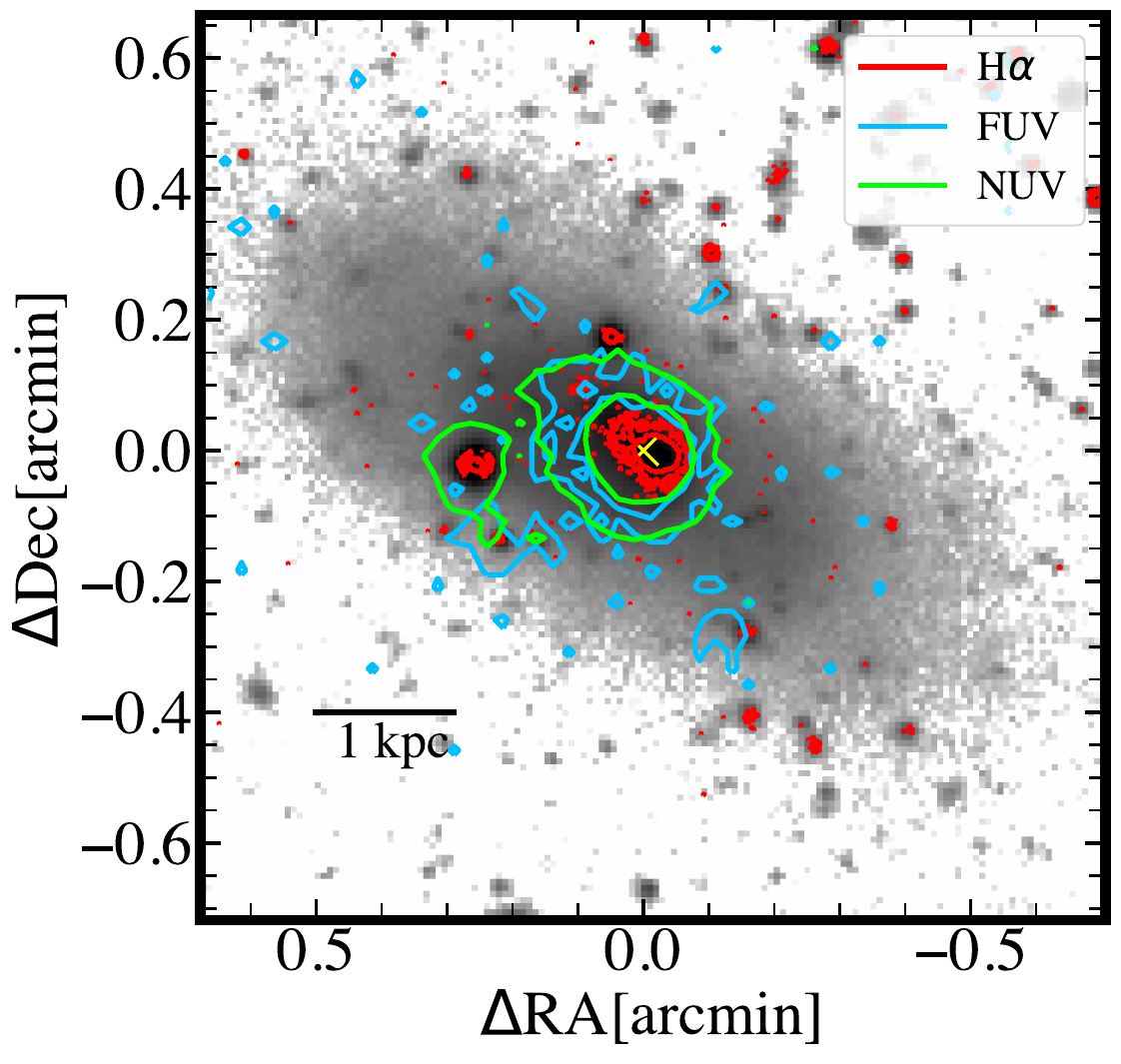}
    \includegraphics[width=0.335\linewidth]{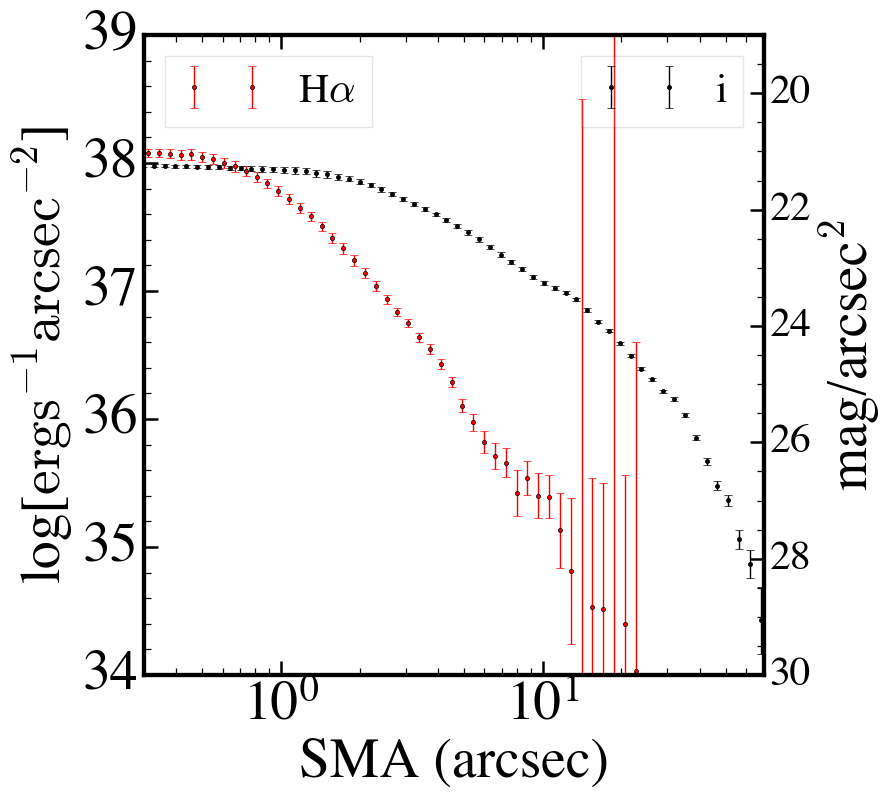}
    \includegraphics[width=0.335\linewidth]{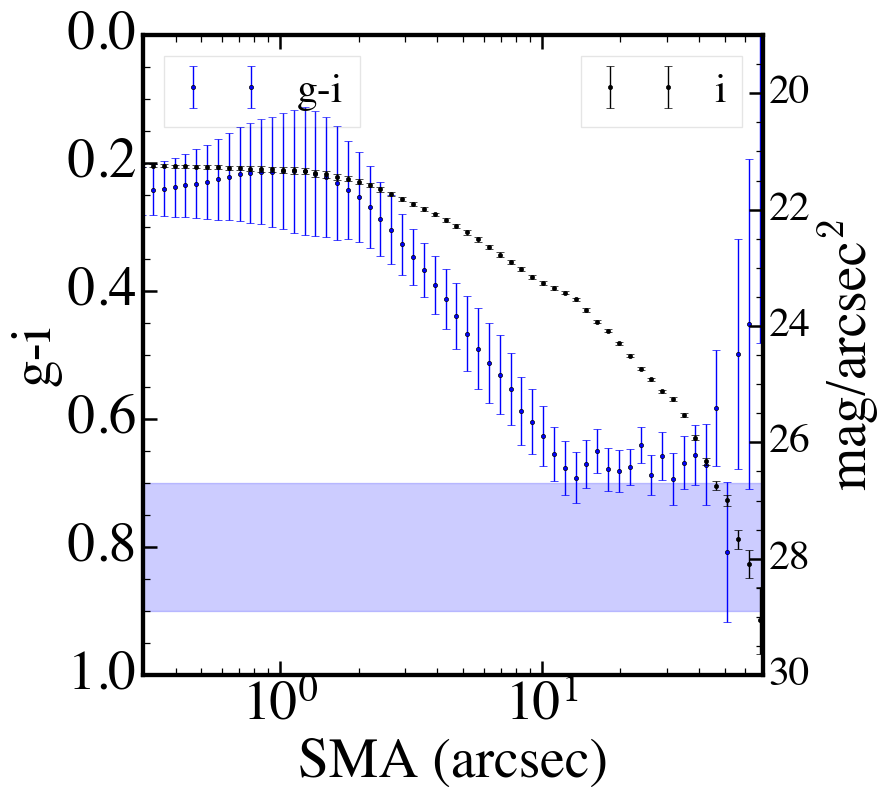}
    \caption{Left panel: Star formation distribution shown by H$\alpha$ (red contour), UV (blue contour for FUV and lime contour for NUV) overplotted in gray-scale NGVS $g$ band image. The contour values are 3 times and 10 times the noise level. The yellow cross marks the galaxy center. Middle panel: Radial profile of H$\alpha$ surface density from the narrow band H$\alpha$ images. We also present the $i$-band surface brightness profile as a reference. Right panel: Color profile of VCC 479 by calculating the difference between $g$ band and $i$ band surface brightness profile (Figure \ref{fig:SB} in Section \ref{sec:SB}). The blue-shaded areas represent the 2$\sigma$ range of the $g-i$ colors of normal dEs at the same magnitudes as VCC 479, derived from the color–magnitude relation of the Virgo cluster core region by NGVS observations \citep{Roediger2017}}
    \label{fig:SFdistribution}
\end{figure*}

We utilized H$\alpha$ and GALEX UV images to trace the recent star formation activity of the whole galaxy. The contours of H$\alpha$, GALEX FUV, and NUV emission are overlaid on the NGVS $g$-band image (left panel of Figure \ref{fig:SFdistribution}). It is obvious that the recent star formation is concentrated in the central region. Additionally, we examined archival Herschel images \citep{Corbelli2012} and found no detection of VCC479 in the Herschel 70$-$500 $\mu$m bands. This indicates a lack of significant dust-obscured star formation. We calculated the SFR traced by FUV and H$\alpha$ (NII contamination have been corrected based on SDSS spectrum) using the calibration provided by \citet{Torrecilla2015}:
\begin{equation}
    \rm SFR(\rm M_\odot \rm yr^{-1})\,=\,4.6\times10^{-44}\times\,L(\rm {FUV}_{corr}),
\end{equation}

\begin{equation}
    \rm SFR(\rm M_\odot \rm yr^{-1})\,=\,5.5\times10^{-42}\times\,L(\rm {H\alpha}_{corr}).
\end{equation}
To correct for internal dust extinction in the FUV and H$\alpha$ flux, we adopted an E(B$-$V) = 0.075 with R$_V$ = 4.05 (Section \ref{sec:SDSS}). Since the star formation of VCC 479 is highly centrally concentrated, the dust extinction derived from the SDSS spectrum in the central 3 arcsec should be largely applicable to the star formation measured from images. By applying the extinction corrections, we derived SFR($\rm H\alpha$) = 0.0026 $\rm M_\odot\,\rm yr^{-1}$, and SFR(FUV) = 0.0039 $\rm M_\odot\,\rm yr^{-1}$. The discrepancy between the two estimates reflects the difference in the SFR when averaged over different timescales, suggesting that the global SFR of VCC479 may have been decreasing over the past 100 Myr or so. The integrated SFR and the stellar mass (Section \ref{SED-resolve}) put VCC 479 about 0.3 dex lower than the star-forming main sequence \citep[e.g., Figure 8 of ][]{Boselli2023}, with a global logarithmic specific SFR of $-$10.32. However, the equivalent width of H$\alpha$, derived consistently from both the SDSS spectrum and VESTIGE narrow-band images in the central 3-arcsec diameter, is $\sim$ 200 \text{\AA}. This indicates that VCC 479 is a dwarf galaxy with a central starburst.

In the middle panel of Figure \ref{fig:SFdistribution}, we show the radial H$\alpha$ surface brightness profile. The \textit{i} band surface brightness profile is overlaid for comparison. The contrast between the H$\alpha$ and $i$-band light profiles underscores that recent star formation in VCC 479 is confined to the central few arcseconds of its extended stellar disk. The centrally concentrated star formation is also evident in the \textit{g}$-$\textit{i} color gradient shown in the right panel of Figure \ref{fig:SFdistribution}. The outer exponential disk has nearly constant color that is close to that of normal dEs, consistent with negligible ongoing star formation. We return to this point through stellar population modeling in Section \ref{SED-resolve}.

\subsection{Structural decomposition of SED and star formation history}\label{SED-resolve}

\subsubsection{SED extraction of the two structural components} \label{sec:SED-extrac}

The above analysis makes it clear that the star formation activity in VCC 479 is highly centrally concentrated and appears to be embedded in a largely quiescent exponential disk with radially invariant red colors. Given the clear spatial separation of young and old stellar populations, it is natural to decompose the multi-band surface profiles into two components and then carry out SED fitting for each of them.  Such a decomposition helps to break parameter degeneracies and bypass the outshining effect, both of which generally compromise the effort of deriving reliable results from SED fitting. The outshining effect refers to the dominance of young, bright stars in the SED, which can obscure the contribution of older, fainter stars, leading to biases in age, mass, and metallicity estimates.

Decomposition of the optical broadband surface brightness profiles is demonstrated in Figure \ref{fig:SB}, where the profiles are decomposed into two components, an exponential disk (red line) and a central light excess (green line). The outer exponential disk corresponds to the underlying old stellar populations, whereas the central component represents the central young stellar populations. A similar method has been employed in \citet{Chhatkuli2023} to estimate the stellar mass of central starburst components in BCDs. The decomposed magnitudes of the two structural components are presented in Table \ref{tab:1}. The shallow imaging depth and low spatial resolution of the GALEX FUV/NUV observations prevent us from obtaining a robust decomposition of the profiles in the UV. However, we learned from the deep u band profile decomposition that the exponential disk of old stellar populations overwhelmingly dominates over the central component beyond the central $\sim$ 12 arcsec in radius, whereas the reverse is true (probably even more so at the shorter FUV/NUV wavelength) within the central $\sim$ 5 arcsec in radius. Therefore, the strategy we adopt for extracting SEDs of the two structural components is as follows. Firstly, we obtained photometry of all wave bands for an elliptical annulus with inner and outer radii of 12 and 20 arcsec (middle panel of Figure \ref{fig:SED-resolve}), based on images smoothed to the spatial resolution of NUV (FWHM $\sim$ 4.9 arcsec). Then, the SED of this elliptical annulus was scaled up to match the integrated i band flux of the old exponential disk derived from the radial profile decomposition. Next, the SED of the central burst component was obtained by subtracting the scaled exponential disk SED from the integrated SED of the entire galaxy (first row in Table \ref{tab:1}, measured by growth curve from the surface photometry in Section \ref{sec:SB}). The resulting SEDs are all given in Table \ref{tab:1}.

\begin{table*}
\caption{UV-optical photometry}
\label{tab:1}
\centering
\resizebox{0.90\textwidth}{!}{
\begin{tabular}{lccccccc}
\hline\hline
{}  & LyC (mJy) & FUV (ABmag) & NUV (ABmag) & u (ABmag) & g (ABmag) & i (ABmag) & z (ABmag) \\
\hline
Galaxy & $0.0037\pm0.0004$ & $19.35\pm0.077$ & $18.90\pm0.056$ & $17.09\pm0.0111$  & $16.30\pm0.0054$ & $15.74\pm0.0071$ & $15.63\pm0.0043$ \\
Outer region & $0.00048\pm0.00007$ & $21.44\pm0.160$ & $21.31\pm0.088$ & $18.73\pm0.0039$ & $17.74\pm0.0014$ & $17.09\pm0.0028$ & $17.00\pm0.0040$ \\
Exponential disk & $0.00141\pm0.00007$ & $20.30\pm0.160$ & $19.81\pm0.088$ & $17.48\pm0.0165$ & $16.43\pm0.0096$ & $15.82\pm0.0119$ & $15.63\pm0.0132$ \\
Central starburst & $0.00232\pm0.0002$ & $19.93\pm0.175$ & $19.53\pm0.119$ & $18.54\pm0.0473$ & $18.27\pm0.0422$ & $18.03\pm0.0797$ & $18.35\pm0.0463$ \\
\hline
\end{tabular}}
\tablefoot{The outer region refers to the region indicated by the elliptical annulus region shown in the left panel of Figure \ref{fig:SED-resolve}. The exponential disk and central starburst refer to the two components from the structure decomposition in Section \ref{sec:SB}.}
\end{table*}

\subsubsection{SED fitting method}\label{sec:SED-method}

We conducted SED fitting using the CIGALE (Code Investigating GALaxy Emission; \citep{Noll2009,Boquien2019}) SED fitting code. We mainly utilized the GALEX FUV/NUV and NGVS optical band images. In order to constrain the most recent star formation ($\sim$ 5 Myr), we converted the H$\alpha$ narrow-band flux to Lyman continuum (LyC) flux, the flux in the Lyman continuum pseudo filter as described in \citep{Boselli2016,Boselli2018a,Boselli2021}. The narrow-band H$\alpha$+[NII] imaging data was corrected for [NII] contamination (10\%) based on the SDSS spectrum. We then utilized synthetic models generated with CIGALE to establish a relation between LyC and H$\alpha$ flux. This calibration is as follows:
\begin{equation}
    \rm LyC[\rm mJy]=\frac{2.1\times10^{-39}\times\,L(\rm H\alpha)\,[\rm erg\,\rm s^{-1}]}{D^2\,[\rm Mpc]}.
\end{equation}

Our SED fitting adopted a \citet{chabrier2003} IMF and employed the stellar population synthesis models of \citet{bc03}, together with the dust attenuation law from \citet{Calzetti2000a}. We incorporated prior information for dust extinction, E(B$-$V) and metallicity derived from the SDSS spectrum. The parameter setting is shown in Table \ref{tab:CIGALEpara}. Although we have photometric data from only seven bands, we make full use of prior information, which—as described below—effectively reduces the number of free model parameters to four.

We used a parametric SFH to fit the outer ellipse region (the second row in Table \ref{tab:1}) and central starburst component (the fourth row in Table \ref{tab:1}).
We adopted a constant SFH to represent the old stellar population formed during a stable, secular evolution phase prior to the merger. Such a form of parametric SFH is consistent with the popular scenario of dE transformation through infall of a late-type galaxy into a cluster \citep{Boselli2008}, following similar methods in the literature \citep{Lisker2008,Kenney2014}. For the young and intermediate populations, we modeled as an exponentially declining (burst) component defined as $\rm SFR(t)\,\sim\,e^{-(age_{\rm burst}-t)/\tau_{\rm burst}}$, where t is the lookback time. Three parameters, $\rm age_{\rm burst}$, $\rm \tau_{\rm burst}$, and the mass fraction of burst component $f_{\rm burst}$ define the parametric SFH. The fitting ranges of these three SFH parameters are indicated in Table \ref{tab:CIGALEpara}.

\begin{table}
\caption{Input parameters for CIGALE}
\label{tab:CIGALEpara}
\centering
\begin{tabular}{lcc}
\hline\hline
Parameter & Value & Units \\
\hline
Pop. synth. model & \citet{bc03} & -- \\
IMF & Chabrier & -- \\
Metallicity & 0.008 & -- \\
E(B-V) & 0.00--0.07 & -- \\
$age_{\text{burst}}$ & 100--6000 & Myr \\
$\tau_{\text{burst}}$ & 10--1000 & Myr \\
$f_{\text{burst}}$  & 0.0--1.0 & -- \\
\hline
\end{tabular}
\end{table}

\subsubsection{Results from SED fitting}\label{sec:SED-result}
\begin{figure*}
    \centering
    \includegraphics[width=0.40\linewidth]{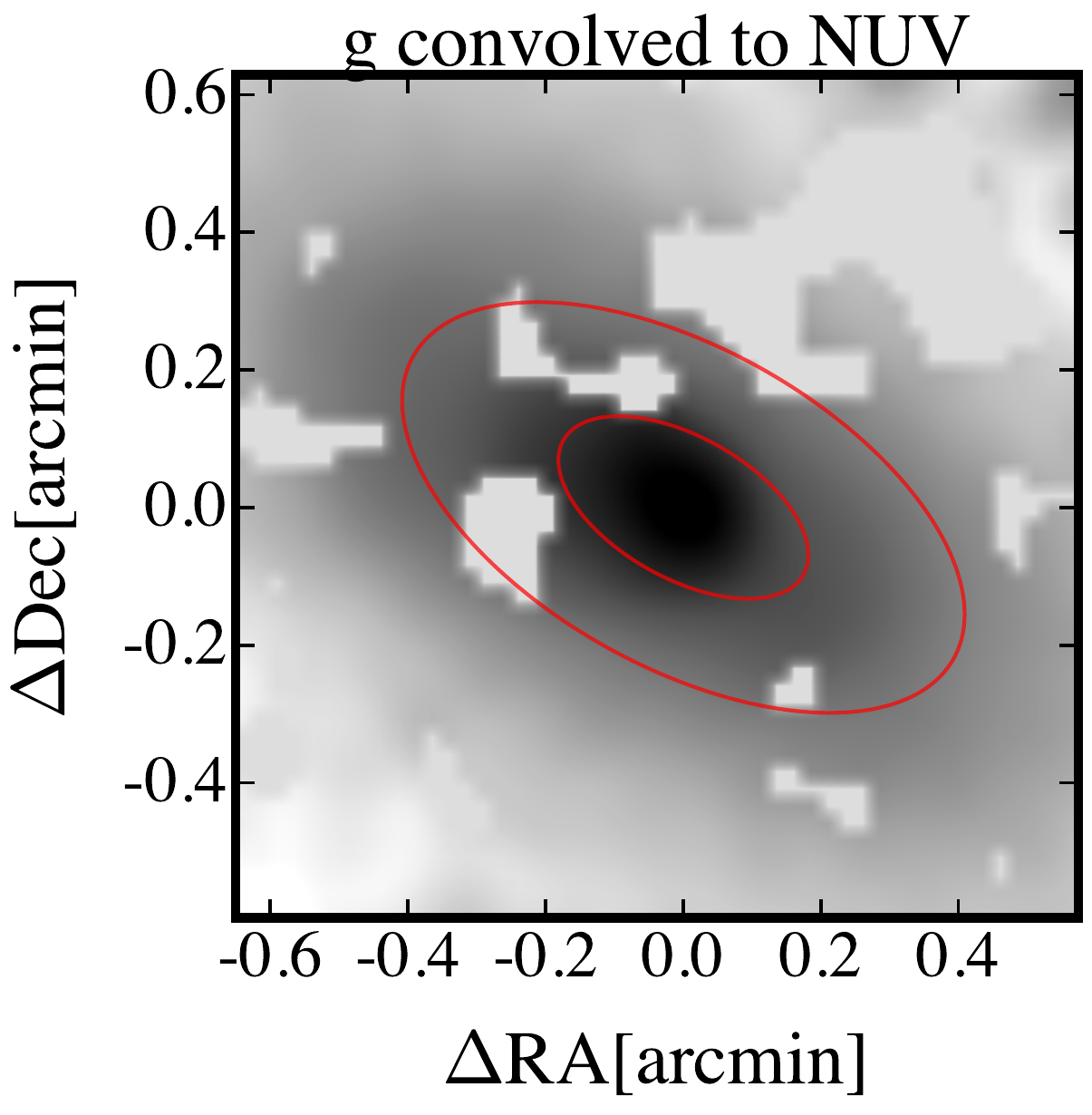}
    \includegraphics[width=0.57\linewidth]{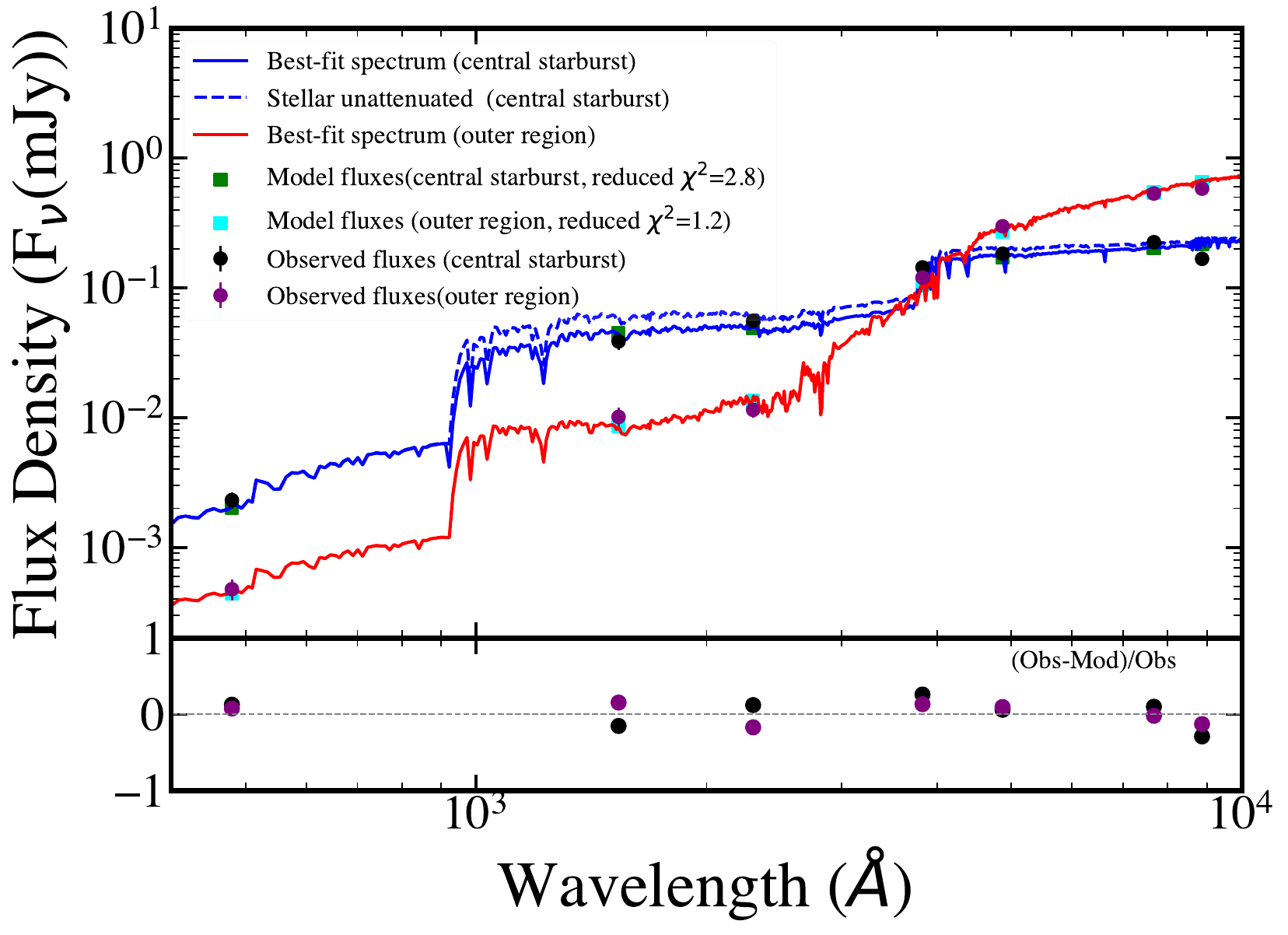}
    \caption{SED fitting of the central blue core and outer quenched disk. Left panel: PSF-matched (to GALEX NUV resolution) $g$ band image. The elliptical annulus in red indicates the region used to extract the SED of the outer disk. Right panel: CIGALE best-fit SEDs of the outer region and the central starburst component. Different colors are indicated in the legend. The residuals are displayed in the lower panel.}
    \label{fig:SED-resolve}
\end{figure*}

\begin{figure*}
    \centering
    \includegraphics[width=0.51\linewidth]{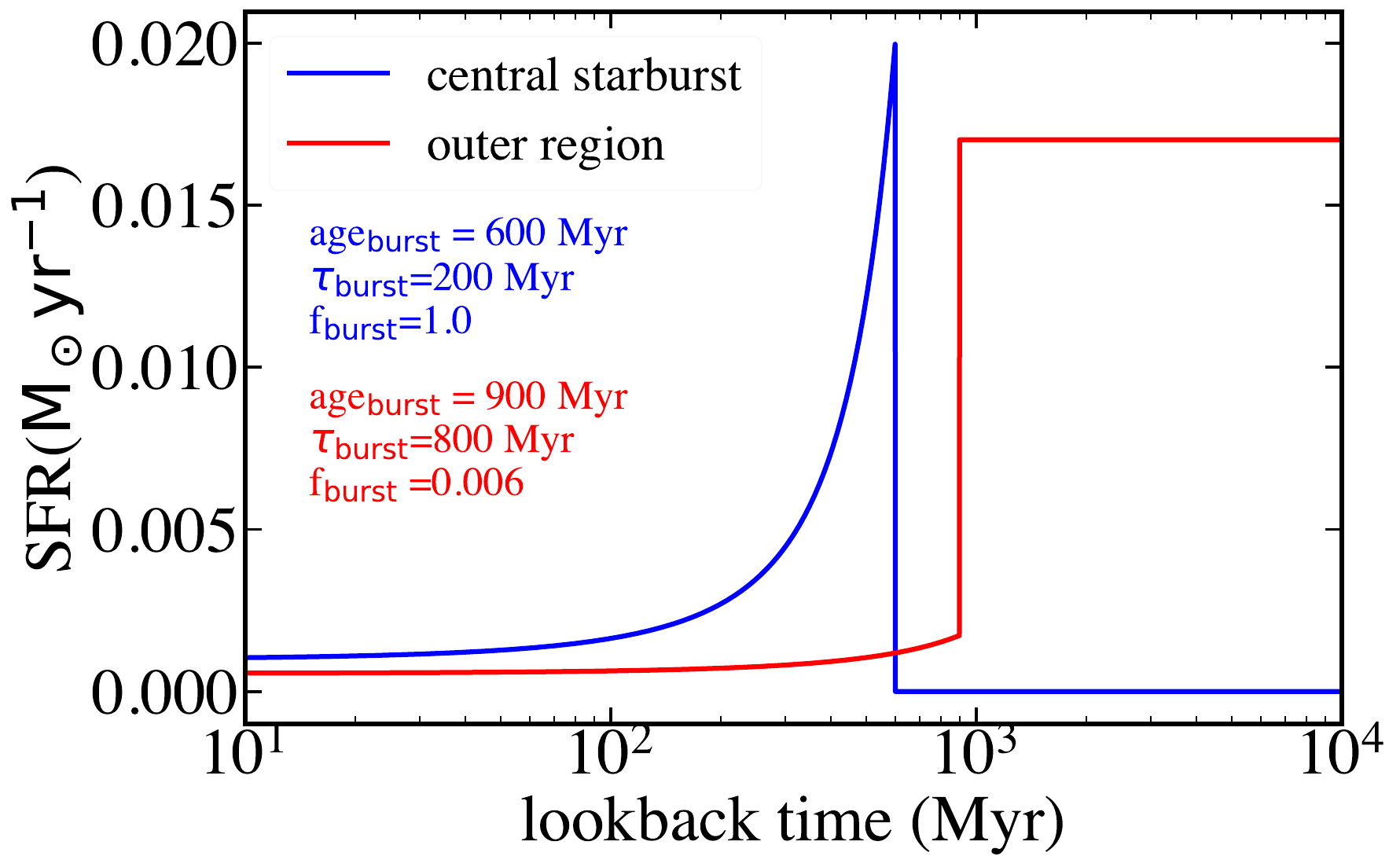}
    \includegraphics[width=0.47\linewidth]{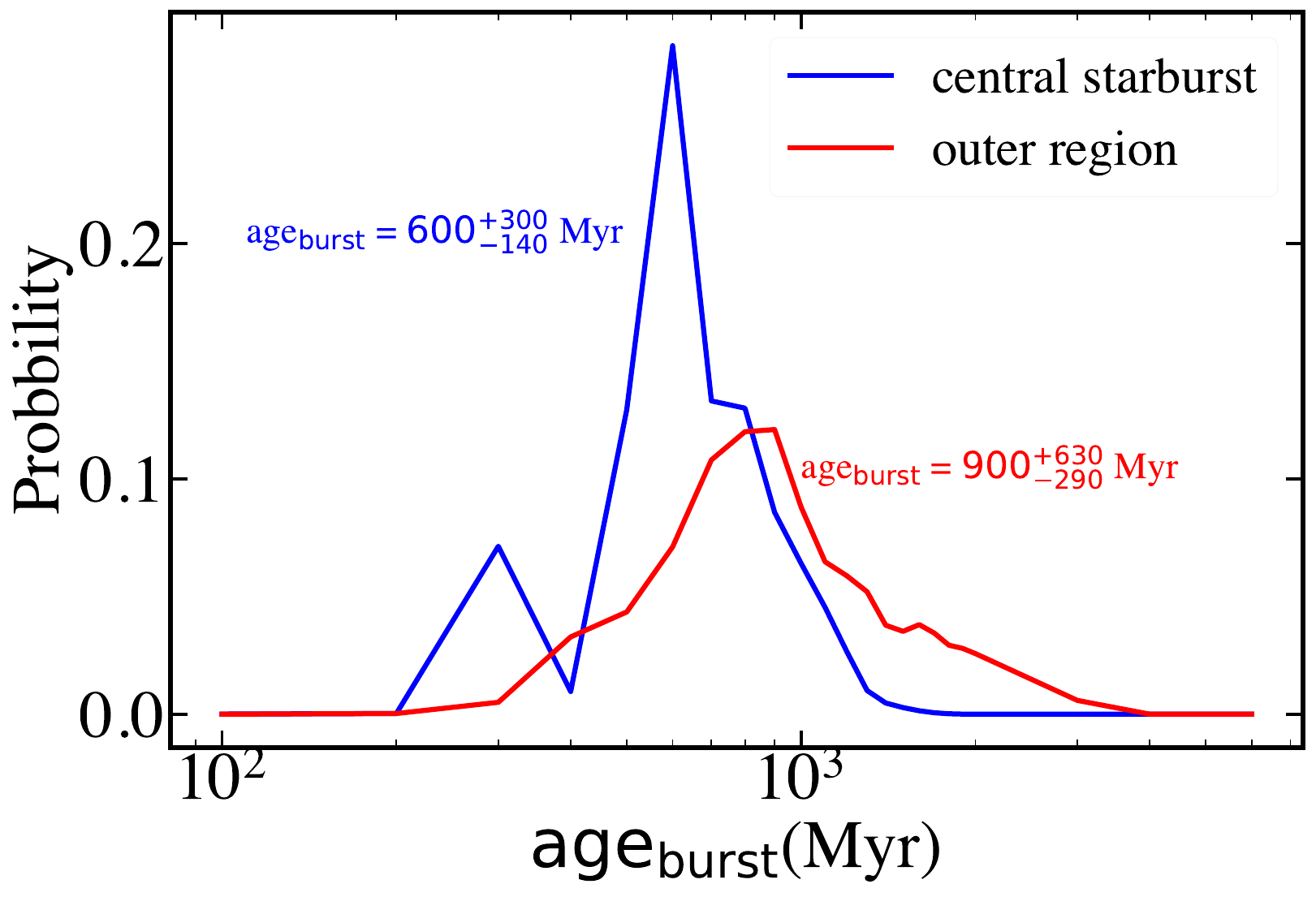}
    \caption{Left panel: Best-fit parametric SFH for the outer quenched exponential disk (red) and central starburst (blue). The best-fit SFH parameters are indicated in the figure. Right panel: Posterior probability distribution of $\rm age_{\rm burst}$}.
    \label{fig:SFH}
\end{figure*}

The best-fit SED of the outer elliptical annulus region (indicated in left panel of Figure \ref{fig:SED-resolve}) is overplotted on the measured data points in the right panel of Figure \ref{fig:SED-resolve}. The SED of the outer region is notably red, with virtually zero dust attenuation. The best-fit parametric SFH and the posterior probability distribution of $\rm age_{\rm burst}$ is shown in Figure \ref{fig:SFH}. We found that the parameter $\rm age_{\rm burst}$ is well constrained. The star formation of the exponential disk was quenched approximately 900$^{+630}_{-290}$ Myr ago. As mentioned above, we scaled up the best-fit SED of the outer elliptical annulus region to match the \textit{i} band flux of the decomposed exponential disk. As expected, the decomposed total flux of the exponential disk by surface brightness profile fitting in the \textit{u}, \textit{g}, \textit{i}, and \textit{z} bands is consistent with the scaled best-fit SED of the outer elliptical annulus region. By scaling up the best-fit SED, we calculated the total stellar mass of the exponential disk component to be approximately $\sim7.8\times10^{7}\,\rm M_\odot$, with an uncertainty of about 8\% as obtained from the CIGALE posterior probability distribution.

The best-fit SED for the central starburst component is also shown in the right panel of Figure \ref{fig:SED-resolve}. The inferred SFH suggests a burst of star formation began approximately 600$^{+300}_{-140}$ Myr ago, with $\tau$=200 Myr. The contribution of the constant-SFR model component in this central structural component is virtually zero. The stellar mass of this starburst component is estimated to be $\sim\,2.4\times\,10^6\,\rm M_\odot$, with an uncertainty of about 10\%. So the total stellar mass of VCC 479, calculated by summing the contributions from the old exponential disk and the central starburst components, is approximately $8.0\times\,10^7\,\rm M_\odot$. The mass fraction of the starburst component with respect to the total stellar mass is about 2.9$\pm0.5$\%, where the error is propagated from that of the stellar mass.

In the upper right corner of Figure \ref{fig:shell}, we show the central 10 arcsec region of the \textit{i}-band image, which has a seeing disk FWHM of 0.53 arcsec. While numerous irregular clumps are visible at the center, no dominating nuclear star cluster was identified. It is possible that these clumps will merge to form a nuclear star cluster in the future. For the bright region marked in red, we used \texttt{photutils} to measure its minor-axis FWHM as 0.9 arcsec. After applying quadratic subtraction of the PSF FWHM, the intrinsic FWHM is 0.52 arcsec, corresponding to approximately 42 pc. We measured the flux of the bright clumps by subtracting the contribution of underlying exponential disk component. By using the stellar mass to light ratio inferred for the central starburst component from SED modeling, the stellar mass of the brightest clumps was calculated to be $3.5\times10^5 \rm M_\odot$.

\section{HI analysis}\label{sec:HI}
\subsection{HI gas distribution}\label{sec:HI-distribution}
\begin{figure*}
    \centering
    \includegraphics[width=0.345\linewidth]{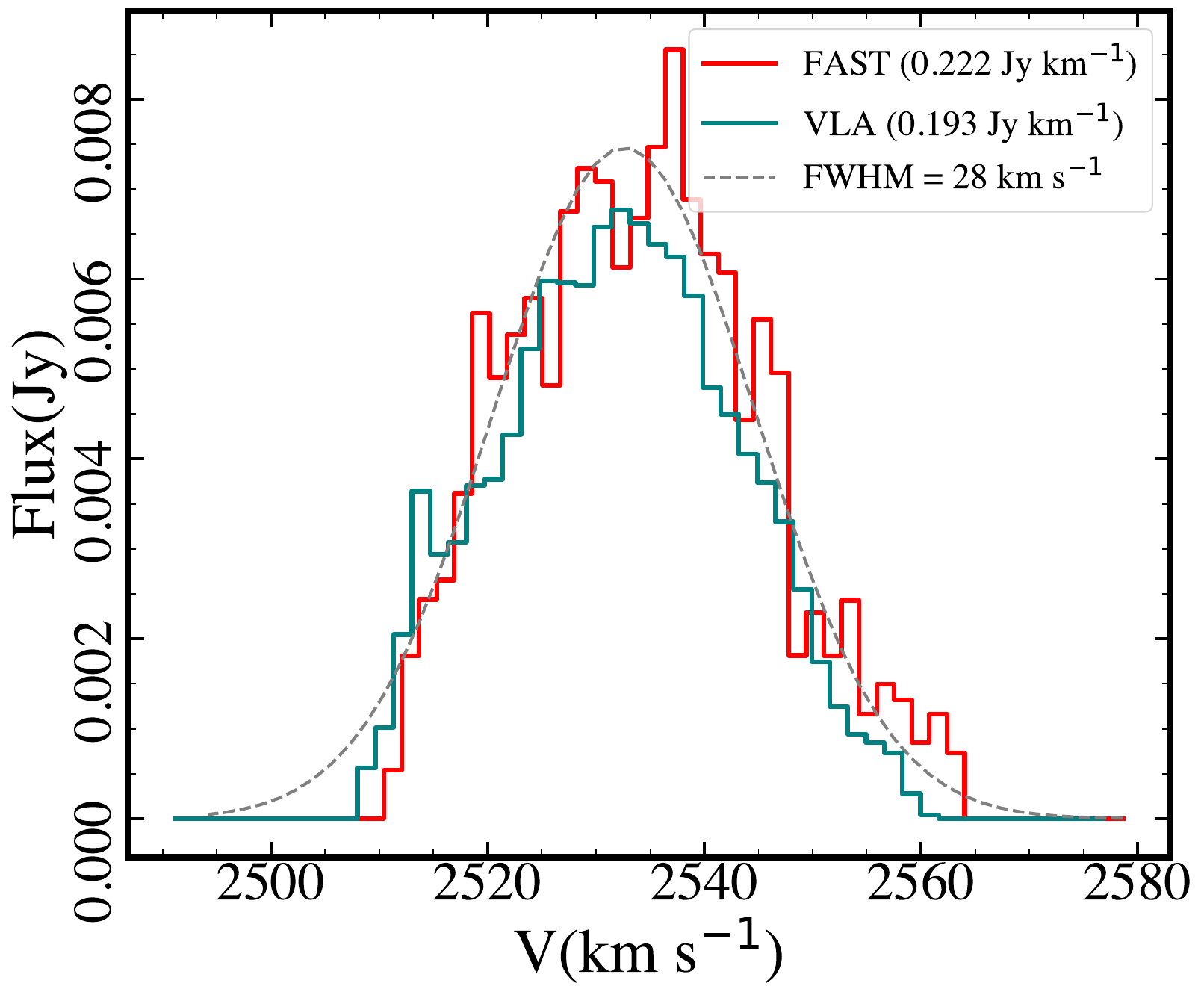}
    \includegraphics[width=0.355\linewidth]{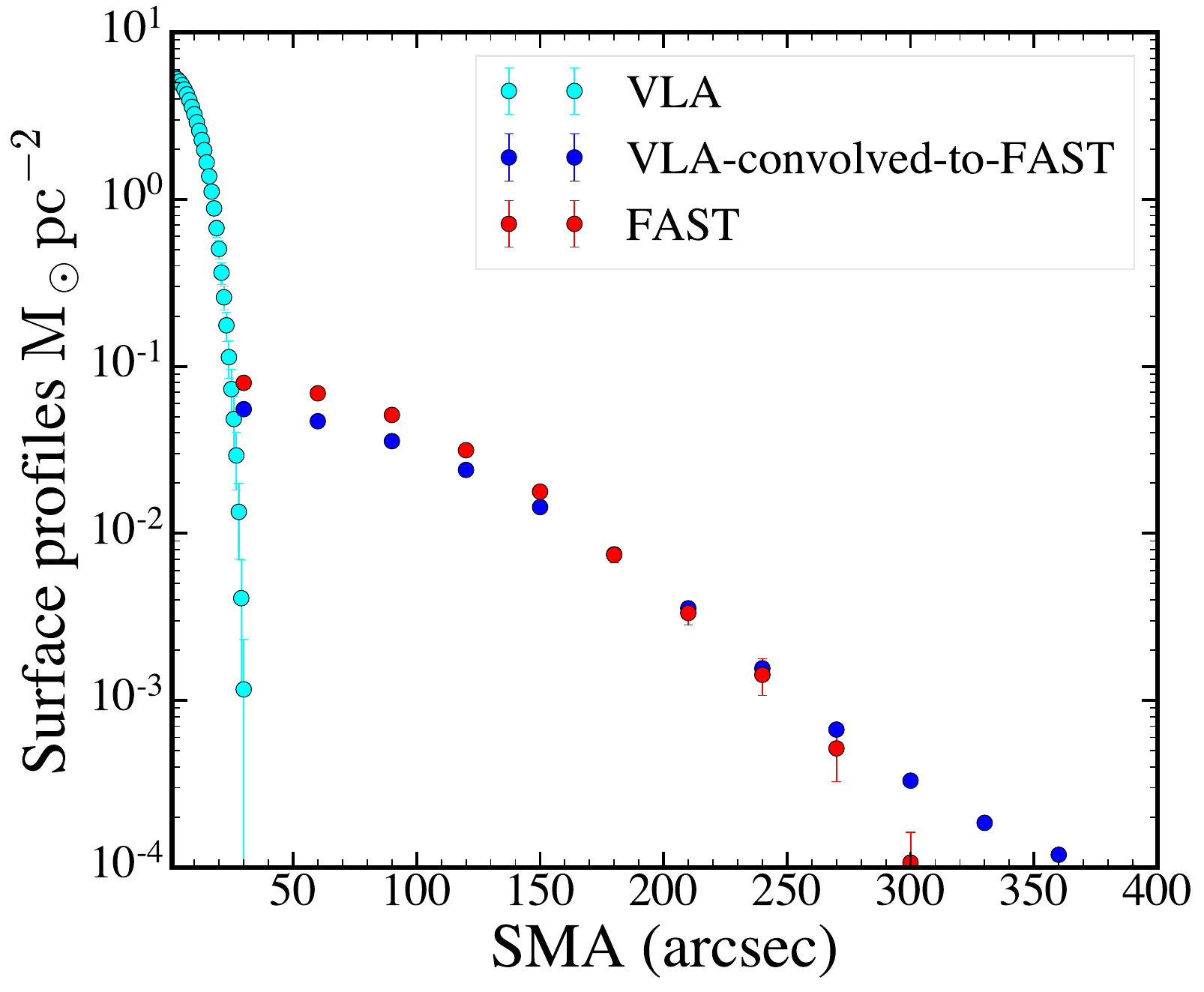}
    \includegraphics[width=0.283\linewidth]{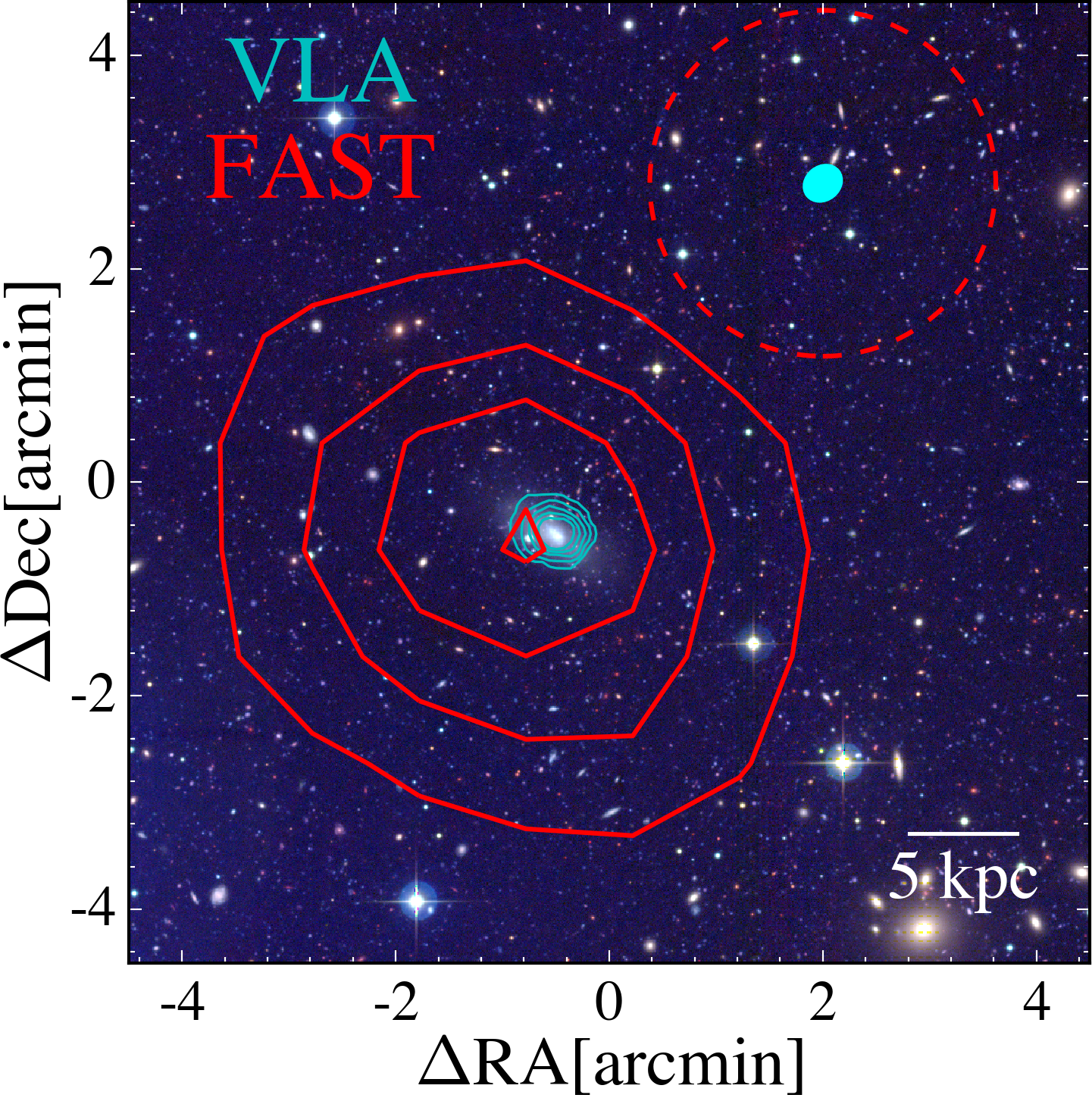}
    \caption{Left panel: Comparison of the FAST and VLA integrated HI spectra. Only pixels with S/N $> 3$ are incorporated in the velocity profiles. The total HI flux densities from the two telescopes are indicated in the legend. The dashed curve shows the best-fit Gaussian profile to the FAST HI spectrum, yielding a FWHM of 28 km s$^{-1}$. Middle panel: Radial profiles of HI column densities derived from VLA and FAST. Radial profile derived from the VLA map that is smoothed to match the FAST beam (FWHM$\sim$3.24 arcmin) is also plotted for comparison.
    Right panel: HI intensity contours of VLA and FAST overplotted on the NGVS color image. The red curves denote the FAST HI (1, 3, 5, 7) $\times\,\rm {10^{18}\,cm^{-2}}$ column density contours. The cyan curves represent the VLA HI (3, 10, 20, 30, 40) $\times\,\rm {7\times10^{18}\,cm^{-2}}$ column density contours. The FAST beam and the VLA synthesized beam are shown in the upper right corner as a dashed red circle and a cyan-filled ellipse, respectively. Note that the tapered natural weighted VLA HI image cube is used here.}
    \label{fig:HI-profile}
\end{figure*}

\begin{figure*}
    \centering
    \includegraphics[width=0.369\linewidth]{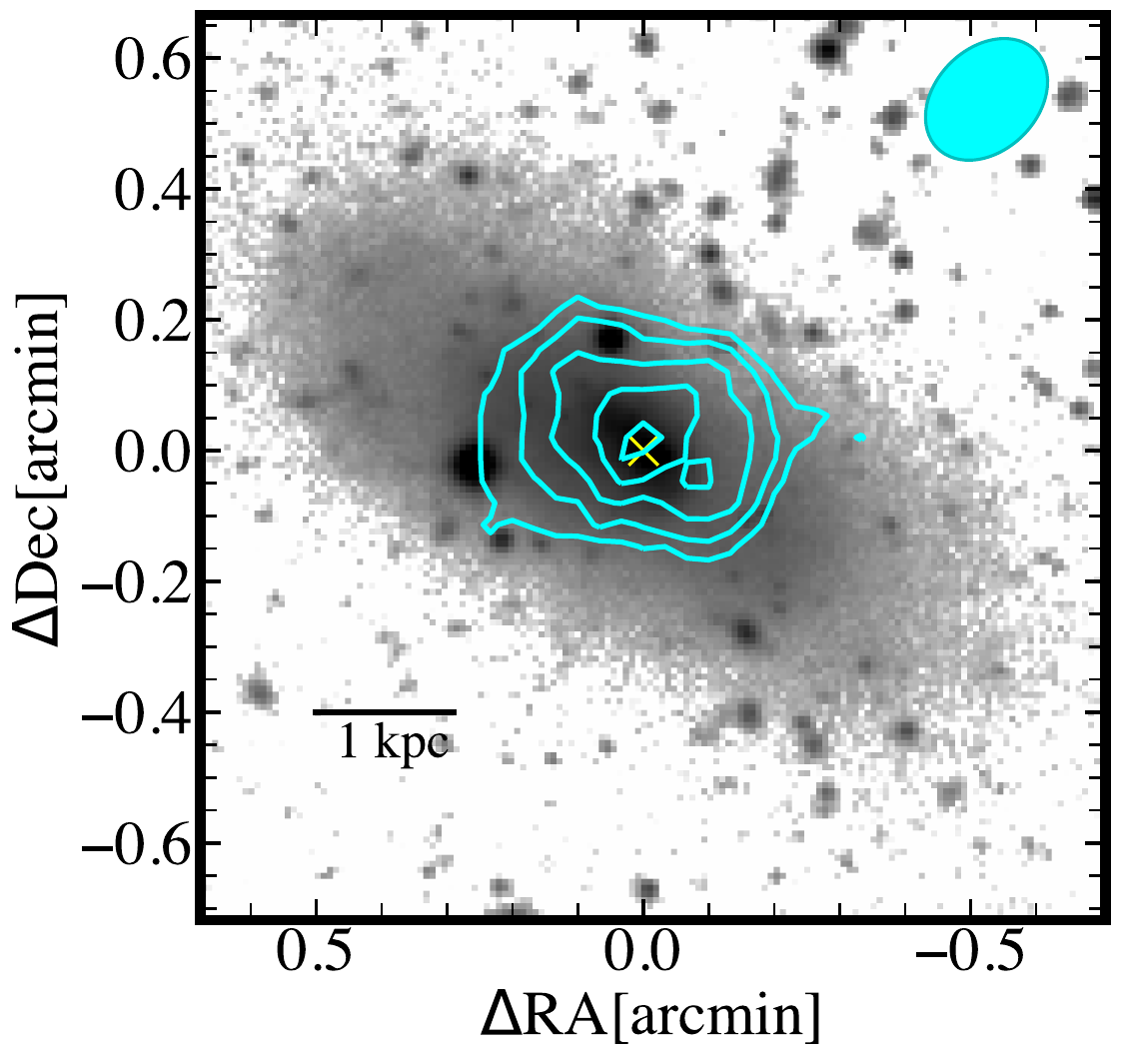}
    \includegraphics[width=0.308\linewidth]{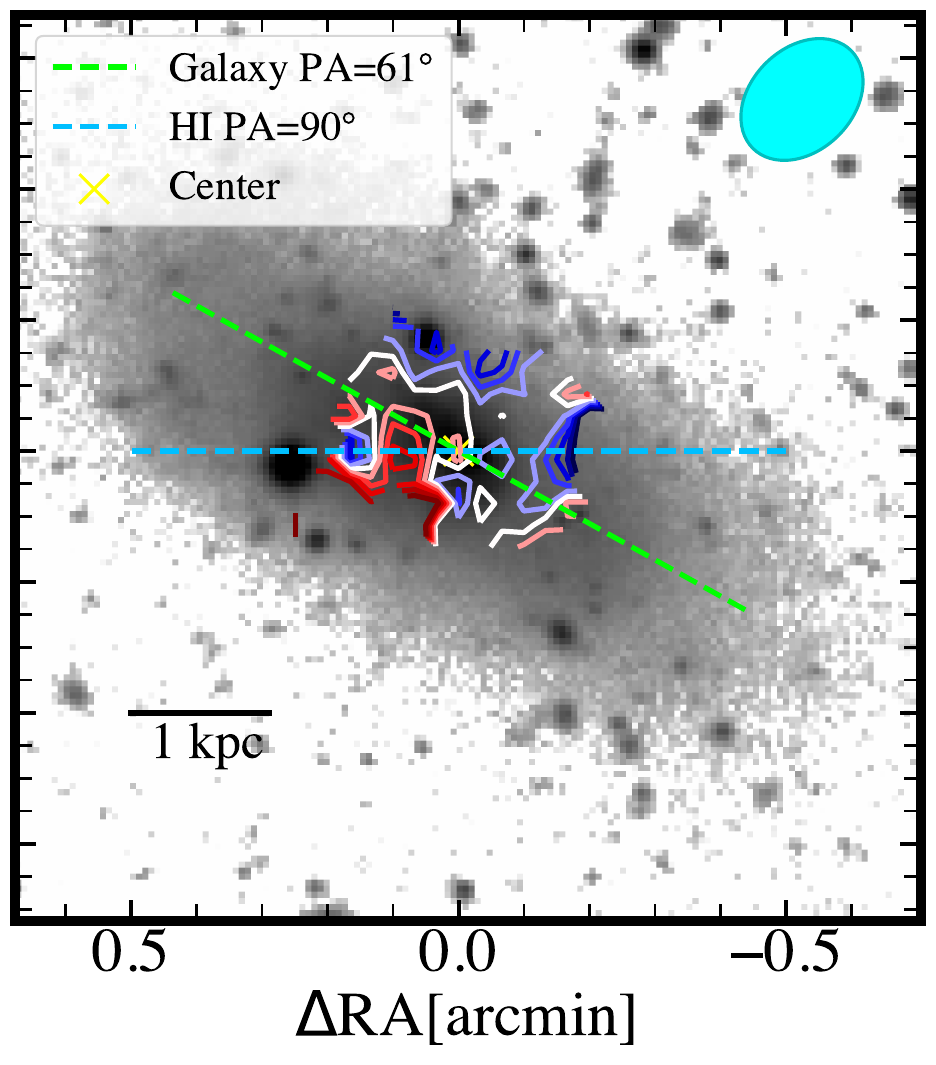}
    \includegraphics[width=0.308\linewidth]{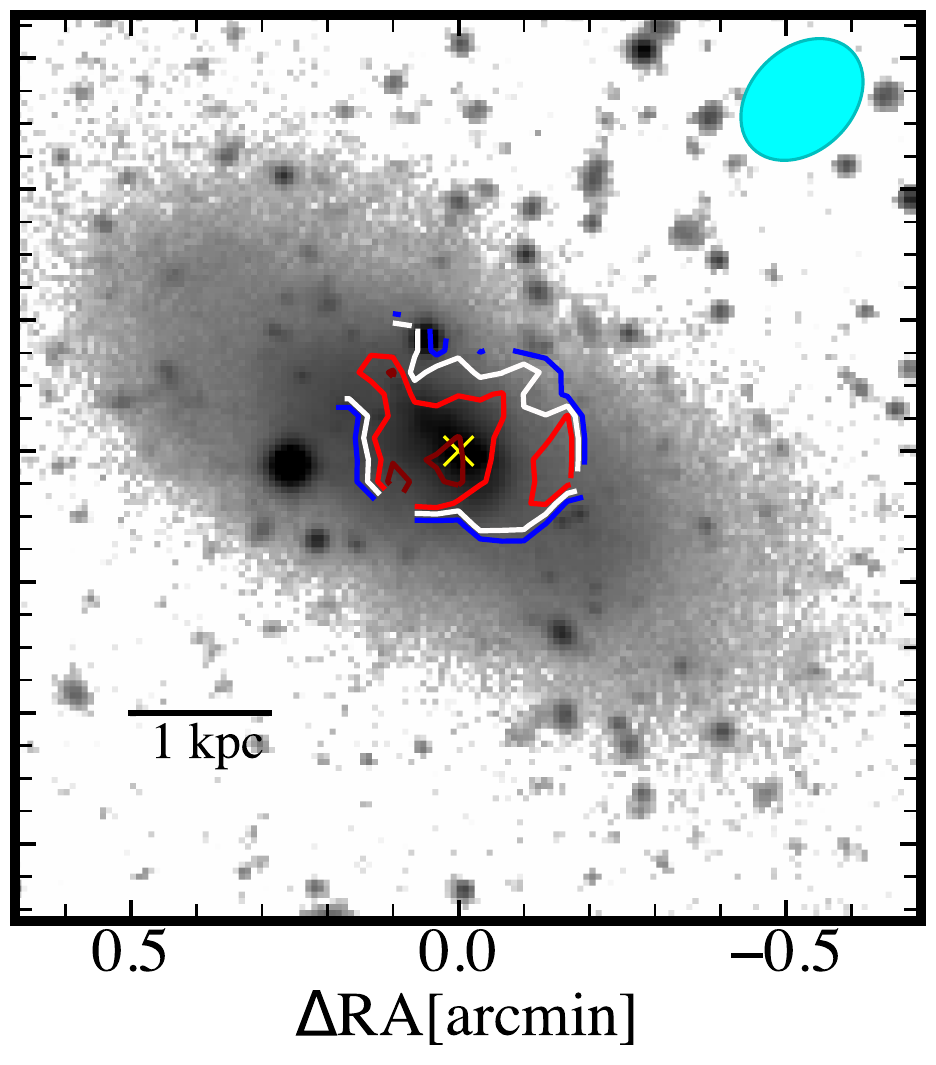}
    \caption{Natural weighted HI moment maps of VLA overlaid on the NGVS $g$-band image. Left panel: Contours of the moment 0 (i.e., intensity) map plotted as cyan curves. The contours are respectively for HI column densities of (3, 10, 20, 30, 40) $\times\,\rm {1.9\times10^{19}\,cm^{-2}}$.  Middle panel: HI velocity field contours drawn at 2 km s$^{-1}$ intervals. The white contour represents the systemic velocity of 2531 km s$^{-1}$. The galaxy's optical and HI major axes are shown as dashed lines in lime (PA = 61\degree) and blue (PA = 90\degree with maximum velocity gradient), respectively. Right panel: HI velocity dispersion contours drawn at 3, 6, 9, 12 km s$^{-1}$. The yellow cross marks the optical center of VCC 479. The synthesized beam is represented by a cyan-filled ellipse in the upper right corner of each panel.}
    \label{fig:HI-map}
\end{figure*}
The HI 21 cm emission provides indispensable and unique insights into tidal interaction and merging processes. The FAST observations, with their high sensitivity, are ideal for estimating the total HI mass. In contrast, the high spatial resolution of the VLA data allows us to map the HI distribution and velocity field in detail. Combining these two datasets allows for a comprehensive understanding of the HI properties of VCC 479.

The integrated HI velocity profiles from the VLA (with tapered natural weighting) and FAST observations are shown in the left panel of Figure \ref{fig:HI-profile}. Only pixels with integrated HI flux densities exceeding three times the RMS noise level are included when constructing the velocity profile. The HI velocity profile is symmetric around the systemic velocity (2531 km $\rm s^{-1}$). The Gaussian fit to the FAST HI velocity profile results in a FWHM of 28 km s$^{-1}$, equivalent to a dispersion of $\simeq$ 12 km s$^{-1}$. The profiles from the VLA and FAST observations are broadly consistent, with the VLA recovering $\simeq$ 87\% of the total HI flux detected by FAST (0.222 Jy $\rm km\,s^{-1}$). The middle panel of Figure \ref{fig:HI-profile} shows the radial profiles of HI mass surface density, respectively, from the VLA and FAST observations, derived using circular annuli. To compare with the FAST data, the VLA HI radial profiles were generated based on HI maps that are convolved to the spatial resolution of the FAST image cube. It can be seen that the resulting VLA HI radial profile broadly agrees with that of FAST. This implies that the VLA observations capture most of the HI emission area and the apparently larger spatial extent of the FAST HI map mainly reflects a larger beam size. The overall HI distribution in VCC 479 is strongly concentrated toward the central 10-20  arcsec in radius.

The total HI gas mass can be calculated using the following equation:
\begin{equation}
    \rm M(HI) = 2.36\times10^5\,\rm M_\odot\,\times\,\rm d^2(\rm {Mpc})\times\,F_{HI}(\rm Jy\,\rm km\,\rm s^{-1}).
\end{equation}
The resulting HI mass is $1.4\times10^7M_\odot$. We estimated the HI deficiency to assess the environmental stripping suffered by VCC 479, such as tidal stripping and ram pressure stripping, as demonstrated by both observations and simulations \citep[e.g.,][]{Bravo-Alfaro2000,Stevens2017}. HI deficiency is defined as the logarithmic difference between the expected HI mass and the observed HI mass:
\begin{equation}
    \rm DEF_{\rm HI}=\rm {log(M_{HI,exp}/[M_\odot])-log(M_{HI,obs}/[M_\odot])}.
\end{equation}

The expected HI mass of VCC479 was derived from the optical size and galaxy morphology using the relation given by \citet{Haynes1984}:
\begin{equation}
    \rm {log(M_{HI,exp}/[M_\odot])\,=\,a\,+\,2b\,\times\,log(\frac{hd_{25}}{[kpc]})-2log(h)},
\end{equation}
where a = 7.45, b = 0.70 are constants corresponding to the morphological-type Im-BCD from Table 3 of \citet{Boselli2009}. $\rm d_{25}$ = 0.68 arcmin is the $B$-band 25th-mag arcsec$^{-2}$ diameter from HyperLeda\footnote{\url{http://atlas.obs-hp.fr/hyperleda/}}, and h = $\rm H_0/(100\,\rm km\,\rm s^{-1})$. Galaxies are generally considered HI-deficient if DEF$_{\rm HI}\,>$ 0.3, which means they contain less than half the expected HI content \citep{Boselli2009}. We find that VCC 479 is highly HI-deficient, with DEF$_{\rm HI}$ = 1.09$\pm0.07$, where the uncertainty is the difference between the type-independent deficiency and the type-dependent deficiency used in \cite{Chung2009}. We discuss possible scenarios that could explain such a high degree of HI deficiency in Section \ref{sec:gaspoor}.

In the right panel of Figure \ref{fig:HI-profile}, we show the NGVS color image of VCC 479 overlaid with contours of the HI column densities from FAST (red) and VLA (cyan). VCC 479 is barely resolved on the FAST HI map, with a slight elongation approximately along the direction of the optical major axis. In contrast, the VLA HI map reveals significant details (left panel of Figure \ref{fig:HI-map}): virtually all HI resolved by VLA is concentrated within the central $\sim$ 1 kpc in radius. This distribution differs markedly from that of typical galaxies, where the HI gas often extends beyond the stellar disk \citep[e.g.,][]{Hunter2012,Kleiner2023}. The central concentration of HI may be attributed to environmental stripping (e.g., tidal or ram pressure) and merger-induced gas inflow from the galaxy's outskirts \citep[e.g.,][]{Hopkins2013,Moreno2021}. Additionally, the major axes of the HI intensity distribution and the weak velocity field (middle panel of Figure \ref{fig:HI-map}) are misaligned with that of the stellar disk, a feature that may not be uncommon in the aftermath of a merger event \citep{Barrera2015,Li2021,Zhou2022}. Notably, the HI gas spatial distribution does not show signs of compression, stretching, or asymmetry. This suggests that VCC 479 is currently not undergoing ram pressure stripping and tidal stripping, which is consistent with its local environment and the large cluster-centric distance at the present time (Figures \ref{fig:AVID-virgo} and \ref{fig:AVID}). However, given its high HI deficiency, it is quite possible that VCC 479 experienced ram pressure stripping or tidal stripping in the past.

\subsection{HI gas kinematics} \label{sec:HI-kin}
\begin{figure*}
    \centering
    \includegraphics[width=0.355\linewidth]{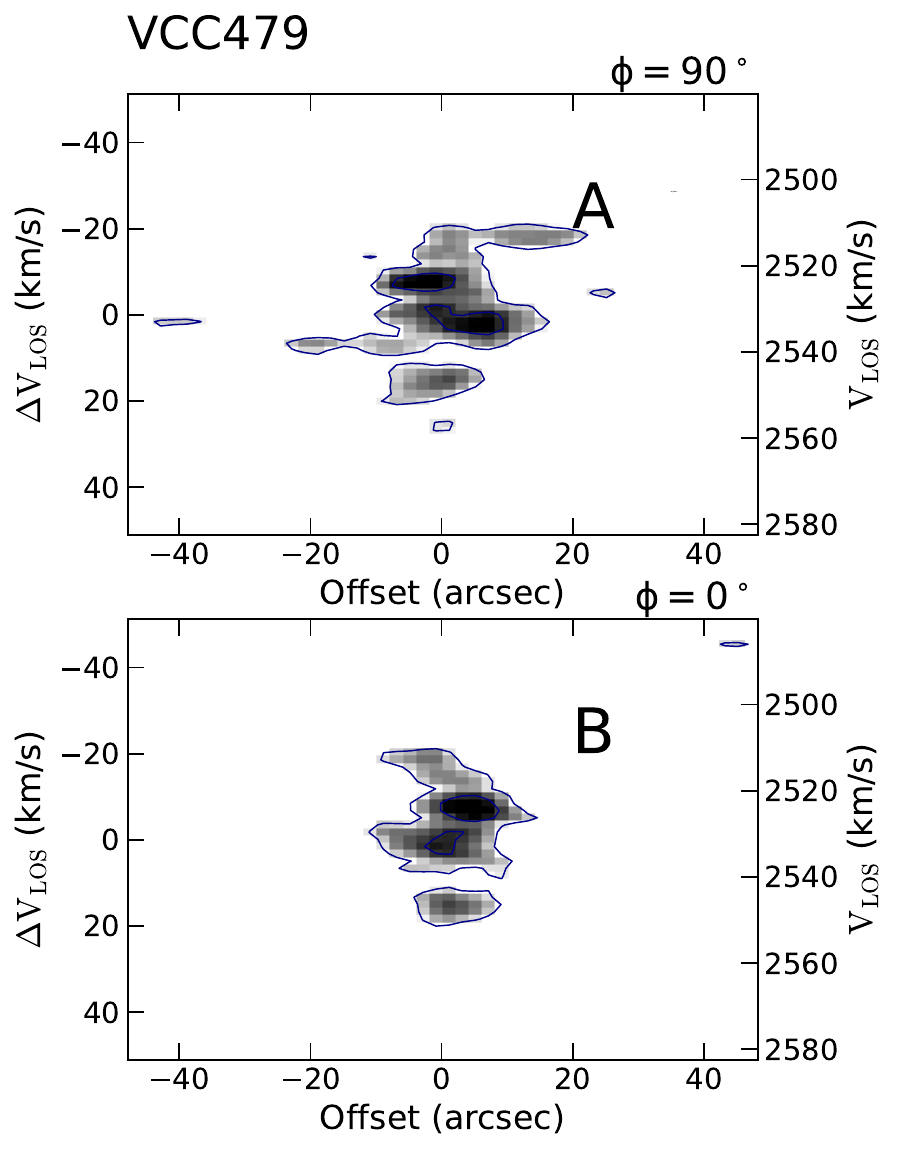}
    \includegraphics[width=0.28\linewidth]{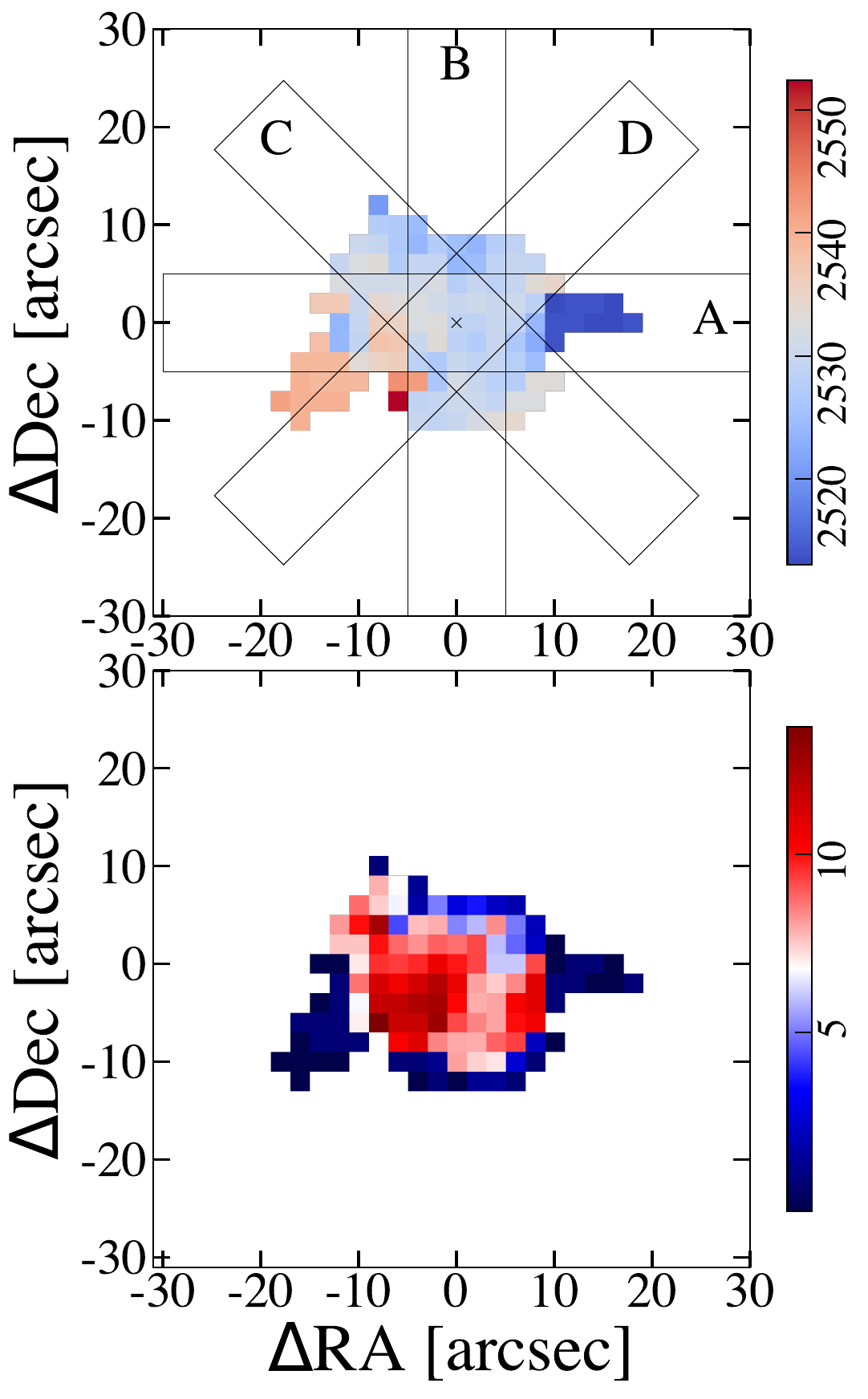}
    \includegraphics[width=0.355\linewidth]{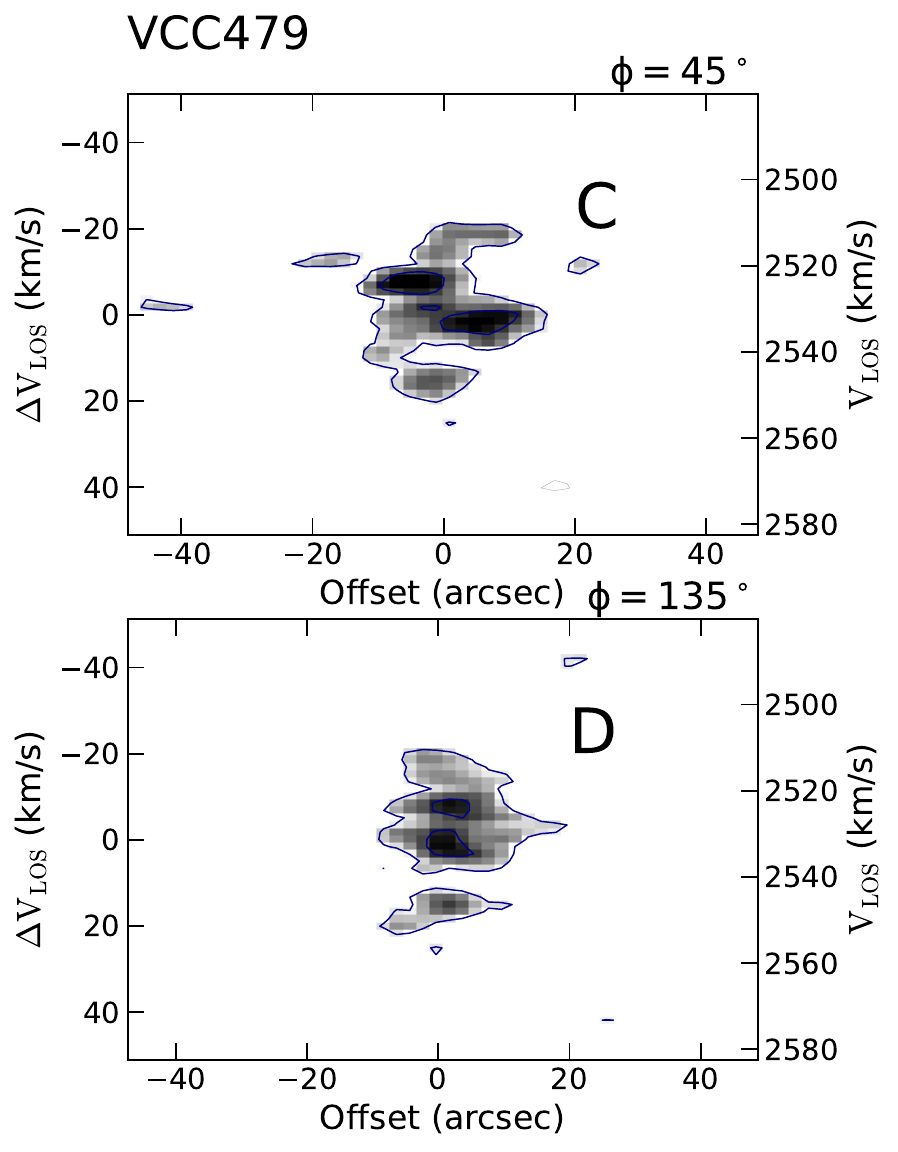}
    \caption{PV diagram of four directions. Upper left panel: PV diagram along the major axis with PA = 90 \degree. Bottom left panel: PV diagram along the minor axis. Upper middle panel: Moment-1 map and directions used to extract PV diagram. The width of the rectangles correspond to the width of the slit used to extract the PV maps. Bottom middle panel: Velocity dispersion map. The color bars are in units of kilometers per second.}
    \label{fig:PV}
\end{figure*}
The HI velocity field, shown in the middle panel of Figure \ref{fig:HI-map} and the upper middle panel of Figure \ref{fig:PV}, displays a shallow velocity gradient that is misaligned (by $\sim$ 30 degree) with the photometric major axis. We attempted to fit a tilted-ring model to the velocity field using the tilted-ring modeling code $^{\rm 3D}${\sc Barolo} \citep{Di2015}. Despite extensive efforts to model the velocity field with a rotating disk, no satisfactory fit can be achieved, indicating non-circular motions. In Figure \ref{fig:PV}, we present the HI position-velocity (PV) diagrams along four directions. The four directions were chosen to be 0 deg, 45 deg, 90 deg, 135 deg from the position angle (90\degree) of the maximum velocity gradient. The PV diagrams reveal chaotic velocity distributions in all directions, without characteristic signs of regular rotation. The moment 2 map shown in the bottom middle panel of Figure \ref{fig:PV} reveals a central velocity dispersion of approximately 12 km s$^{-1}$—higher than the thermal velocity dispersion of warm HI gas ($\sim$8 km s$^{-1}$), yet typical for late-type dwarf galaxies \citep[e.g.,][]{Elmegreen2022}. Therefore, the exceptionally centralized starburst activity appears to have little impact on the HI gas kinematics.

\section{Simulation of a damp dwarf-dwarf major merger}\label{sec:simulation}

To complement the observational results, we carried out simulations of mergers between gas-bearing dwarfs.
While there exist many comprehensive numerical studies of giant galaxies merging, there are significantly fewer studies of the same in the dwarf regime. The usually gas-rich nature and shallow gravitational potential well of dwarf galaxies means the general outcome of mergers can be very different from that of mergers of giant galaxies.

We performed our controlled numerical simulations with the adaptive mesh refinement code Ramses \citep{Teyssier2002}. Stable initial conditions of our models were built using DICE \citep{Perret2014}, and consist of NFW dark matter halos \citep{Navarro1996}, and exponential disks of stars and gas.

\begin{figure}
    \centering
    \includegraphics[width=\columnwidth]{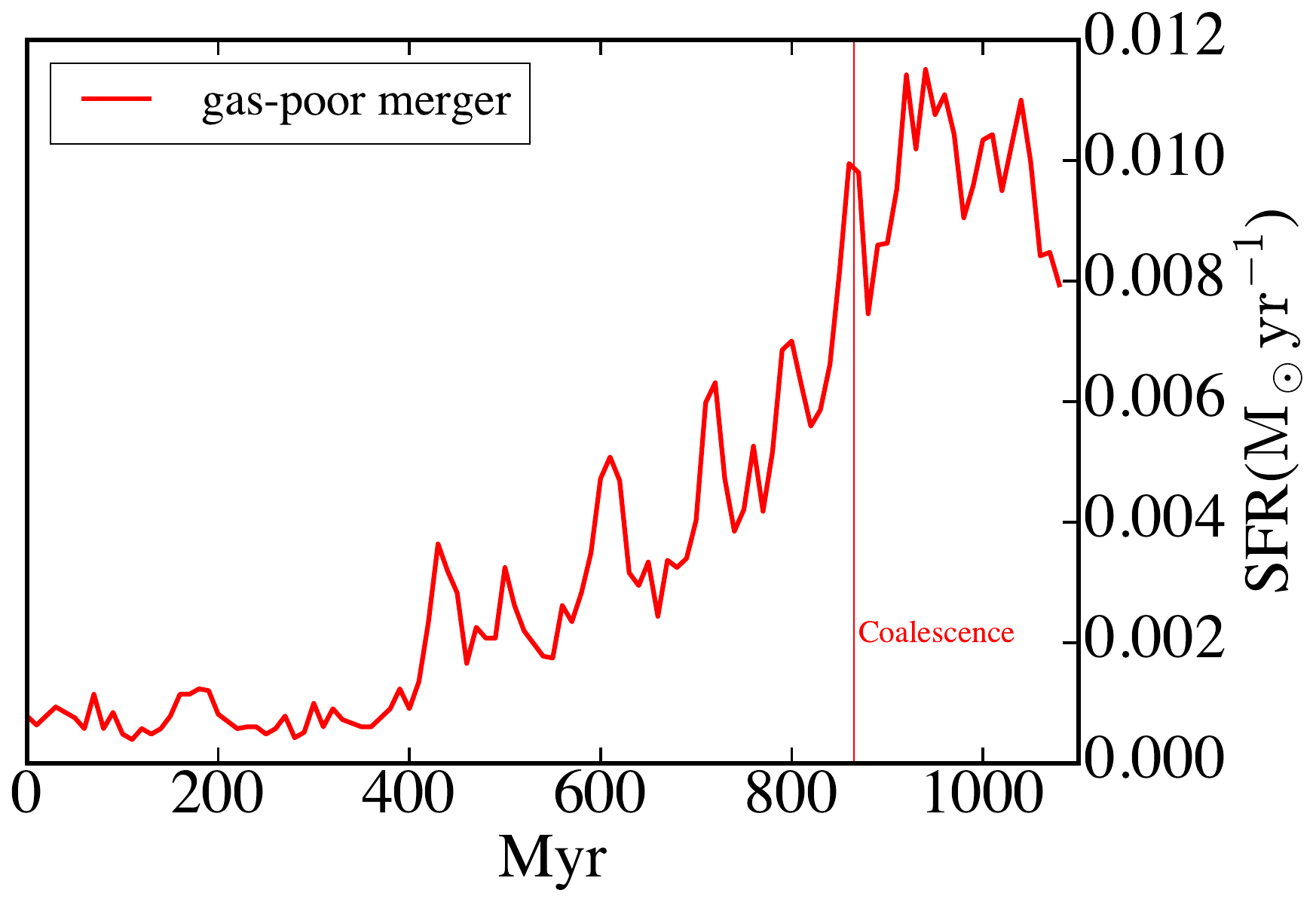}
    \includegraphics[width=\columnwidth]{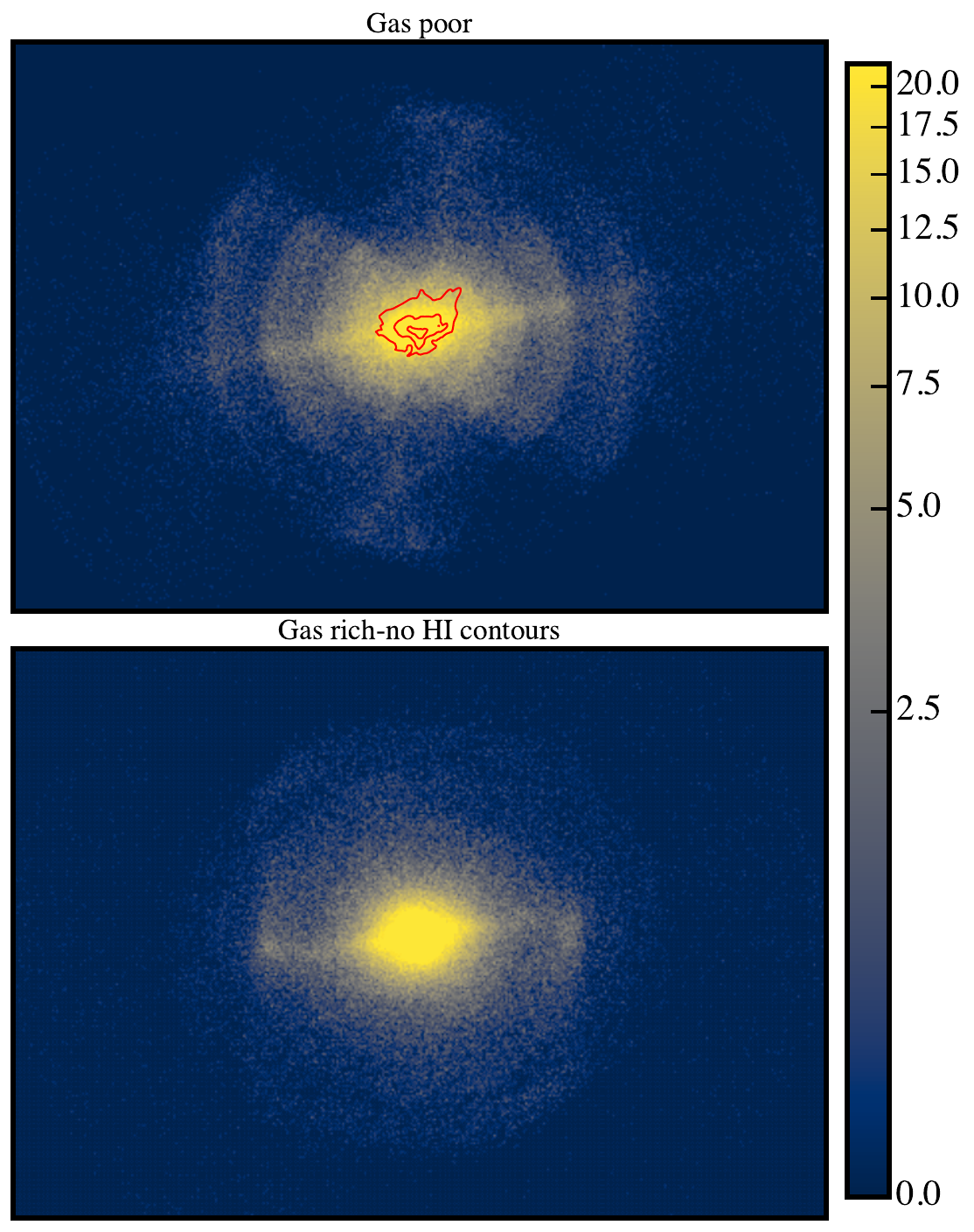}
    \caption{Upper panel: SFR evolution of gas-poor major merger analogous to VCC 479. Middle panel: Stellar particle map overlaid with HI gas contours at 0.5, 5, and 50 $\rm M_\odot\,\rm pc^{-2}$ for the gas-poor merger. The stellar map is expressed in units of solar masses per square parsec. Bottom panel: Stellar map of the gas-rich merger.}
    \label{fig:simulation}
\end{figure}

In this study, we consider the case of an equal mass (1:1 mass ratio) merger between two dwarfs. We use the same model galaxy for both dwarfs involved in the merger, with a dark matter halo with M$_{200}$=6$\times10^{10}\,\rm M_\odot$ and a halo concentration of 7. The stellar disk mass, chosen to be in agreement with halo abundance matching \citep{Behroozi2014} is M$_\star$=1$\times10^{8}\,\rm M_\odot$ with an exponential scalelength of 1 kpc. We chose for the disk to be thickened, as expected for dwarf galaxies of this mass \citep{Sanchez-Janssen2010}, with a vertical to radial scale ratio of 0.3. Our fiducial model (shown in middle panel of Figure \ref{fig:simulation}) has an initial gas fraction of 0.1 (gas mass $\sim$ 1$\times10^{7}\,\rm M_\odot$), representing the scenario of substantial gas loss prior to the merger. We also consider a model with a gas fraction of 1.0 (gas mass $\sim$ 1$\times10^{8}\,\rm M_\odot$), a wet merger, in  order to understand the importance of the gas fraction on the final result.

We chose a box size of $\sim$ 100 kpc which is sufficient to easily contain the two model dwarf galaxy disks and their dark matter out to $>$ 0.5 R$_{200}$. The two galaxy models are placed 10 kpc apart along the x axis. Their disks rotate in the x-y plane, and their relative velocity along the x axis is 100 km s$^{-1}$. Instances of such low relative velocities have been observed in dwarf–dwarf galaxy pairs \citep[e.g.,][]{Stierwalt2015}. To avoid having a fully direct collision, we gave both disks an additional 15 km s$^{-1}$ initial velocity in the positive and negative z direction (out of the plane of the disks). As part of the full simulation dataset, we also considered alternative merging configurations, but in this study we focus on this fiducial configuration. All simulations were conducted for 900 Myr with a snapshot interval of 14.9 Myr. This is sufficient time for the merger to fully coalesce.

The minimum refinement level is six and the maximum refinement level is 11, meaning a minimum cell size of 48 pc. Adaptive mesh refinement is controlled by multiple criteria, including the number of dark matter and star particles from the initial conditions that are found in a cell (we choose 50 for all), and the baryonic (stars and gas) density in a cell which depends on the refinement level of the cell. Additionally, refinement occurs to ensure the Jeans length is resolved to a factor of four, up until the maximum refinement level. This choice of refinement criterion ensures that the gas and stellar disk is resolved to the maximum level of refinement out to the edge of the disk and also vertically out of the plane of the disk.

Our star formation and feedback recipe is based on recipes implemented in \citet{Perret2014}. Star formation occurs for gas above a density threshold of 100 atom$\,\rm cm^{-3}$, and follows a Schmidt relation dependency on gas density with a star formation efficiency of 0.01. Supernova feedback from OB stars is treated as follows. The OB type stellar populations that reach an age of 10 Myr are transformed into supernovae, and release energy, mass and metals into the nearest gas cell. The energy release is 10$^{51}$ erg per 10 M$_\odot$ supernovae, and the release is treated kinetically. 20\% of the mass of a star particle that goes supernova is returned to the gas with a yield of 0.1. For the gas-rich model, this choice of SF parameters reproduces the expected SFR of 0.01 M$_\odot\,\rm yr^{-1}$ for a galaxy of this mass according to the SF main sequence. Gas cooling occurs radiatively \citep{Sutherland1993}, assuming a gas metallicity of 0.15, down to a temperature floor. The temperature floor is dictated by the Truelove condition \citep{Truelove1997} to guard against artificial fragmentation of the gas.

We find that the gas-poor merger simulation generally matches the observations of VCC 479 remarkably well in terms of shell structures, gas distribution, and the fraction of newly formed stars. In Figure \ref{fig:simulation}, we present the evolution of the SFR (top) and the stellar and gas distributions for the gas-poor merger (middle) in the snapshot representing the end of the simulation. A comparison of the shell structures in gas-poor and gas-rich models (bottom panel) highlights the more prominent shells in the gas-poor case. The presence of gas may have altered the dynamics of the merger, thereby suppressing or modifying the formation of prominent stellar shells. This apparent suppression of shell features in the gas-rich simulation may offer further support for a gas-poor merger scenario for VCC 479. A more systematic exploration of this phenomenon, including variations in orbital configurations, will be presented in a follow-up paper (Smith et al. in prep). In the gas-poor merger, the gas is concentrated at the center of the galaxy due to the gas inflow. We can clearly see the enhancement of the SFR. However, the newly formed stars contribute only about 2\% of the final stellar mass ($\sim$4.2$\times$10$^{6}\,\rm M_\odot$), due to the lack of gas. For the gas-rich merger, the mass of newly formed stars during the merger is comparable to that of the pre-existing stellar population. Similarly, simulations of dwarf-dwarf mergers by \citet{Bekki2008} found that gas-rich merger remnants are dominated by newly formed stars and evolve into BCDs.

\section{Discussion} \label{sec:discussion}

\subsection{Origin of the high HI deficiency of VCC 479}\label{sec:gaspoor}
VCC 479 shows an HI deficiency of Def$_{\rm HI}$ = 1.09, indicating that the HI content is an order of magnitude lower than that of normal late-type dwarfs. Typically, late-type dwarf galaxies have HI masses equal to or greater than their stellar masses, as shown by the HI-stellar mass relation \citep{Maddox2015}. We investigate how the HI content of VCC 479 has evolved from a normal late-type dwarf galaxy to its current state.

Previous studies generally suggested that isolated dwarf galaxies do not quench through secular processes \citep{Geha2012}. The quenching of dwarfs is primarily attributed to environmental effects, such as ram pressure stripping or tidal interactions in cluster and group environments. However,
\citet{Koppen2018} calculated the local ram pressure experienced by VCC 479 to be effectively zero, classifying it as a galaxy that underwent stripping in the past but is not currently experiencing ram pressure stripping. The nearby environment is clearly illustrated in Figure \ref{fig:AVID-virgo}, where VCC 479 is located outside the outermost diffuse X-ray emission contours of Virgo. The symmetric HI intensity distribution of VCC 479, showing no sign of compression or tails (Figure \ref{fig:HI-map}), confirms the absence of ongoing ram pressure stripping.

The relatively low mass contribution ($\sim$3\%) by the central starburst aligns with the that expected for a damp major merger (Section~\ref{sec:simulation}). Our simulation in Section~\ref{sec:simulation} suggests that an order of magnitude greater mass contribution by merger-triggered starburst would be expected for a gas-rich major merger \citep[see also observational results of a gas-rich major merger by ][]{Zhang2020a}. Therefore, most of the gas in the progenitor galaxies of VCC 479 may have been lost prior to the coalescence phase of the merger.

Given the projected location of VCC 479 in the phase-space diagram (right panel of Figure \ref{fig:AVID}), we can infer the time of its first crossing (if any) of the virial radius of Virgo cluster in a statistical sense to be $\sim$ 2 Gyr ago \citep{Pasquali2019}.
If we consider the Virgo B subcluster (M49 as the central galaxy) closest to VCC 479 in projection, the minimum crossing time estimated by dividing their projected separation (2.06$\degree$) by the velocity dispersion of the subcluster \citep[464 km s$^{-1}$,][]{Boselli2014} is approximately 1.3 Gyr.
The pre-merger interacting pair of VCC 479 is unlikely to have passed through the cluster center (M87), as the cluster’s tidal force could have exceeded the pair’s gravitational binding force and prevented the merger from occurring. It is plausible that VCC 479 has passed through the outer intracluster medium of Virgo in the past 1–2 Gyr, during which the HI content of the two interacting progenitors experienced substantial ram pressure stripping, and by the time of merging the system has become relatively gas-poor. The relatively gas-poor nature before coalescence also aligns with the quenching timescale of the underlying disk ($\sim$ 900$^{+630}_{-290}$ Myr) inferred from SED modeling. We consider ram pressure stripping to be the most likely mechanism for gas loss of VCC 479. Because we identify a truncated HI gas disk (Figure \ref{fig:HI-map}) with respect to an extended stellar disk. Other hydrodynamical effects, such as starvation, can be ruled out \citep[e.g.,][]{Boselli2022}.

\subsection{Major merger of dwarf galaxies in a dense environment}\label{sec:merger}
The observed features of VCC 479, including shell structures, gas concentration, and modest starburst, can be qualitatively reproduced by simulations of a damp major merger. VCC 479 was selected as a typical dwarf merger candidate from the AVID sample based on the prominent shell structures visible in the NGVS images (Figure \ref{fig:shell} and Figure \ref{fig:slit-profile}). The symmetrical shell structures, together with the similar colors of shells to the stellar main body (middle panel of Figure \ref{fig:slit-profile}, Section \ref{sec:shell}), strongly suggests a major merger with progenitors of comparable masses \citep{Paudel2017}. The resolved HI gas distribution and velocity field have a major axis misaligned with the stellar main body (Figure \ref{fig:HI-map}), which is a characteristic often associated with mergers \citep{Barrera2015,Li2021,Zhou2022}. Furthermore, the HI is highly concentrated in the central region of VCC 479. Whereas, HI in ordinary dwarf galaxies extends beyond the stellar disk.
The extreme central concentration of HI in VCC 479 is likely a result of gas inflow triggered by the major merger.

The prominence of the shell features indicates a recent merger, with the final coalescence occurring about several hundred million years ago \citep{Paudel2017}. According to our stellar population analysis,  the central starburst probably commenced in the past 600$^{+300}_{-140}$ Myr or so. Star formation in the outer exponential disk was quenched about 900$^{+630}_{-290}$ Myr ago, which should correspond to the epoch when the majority ($\sim$90\%) of the gas in the progenitor pair of dwarfs was stripped by environmental effects (as discussed in Section \ref{sec:gaspoor}). With a comparable stellar and HI mass, the progenitor galaxies of VCC 479 likely had similar morphological types. Our simulations also indicate that damp gas-poor major mergers are more likely to produce prominent shell structures compared to gas-rich mergers (Figure \ref{fig:simulation}), which lends further support to the gas-poor nature of the merger event.

Our analysis reveals the distinctive outcomes of major mergers of dwarf galaxies in dense environments. Dwarf galaxies involved in tidal interactions are found to have more extended HI gas distribution than the isolated ones \citep{Pearson2016}. This means that interacting dwarfs are more susceptible to environmental stripping effects during their early stages of merger. The high HI deficiency of VCC 479 can thus be well explained by severe environmental stripping of gas before the final coalescence (Section \ref{sec:gaspoor}). The major epoch of stripping happened about 1 Gyr ago, as indicated by the quenching time of the outer exponential disk of VCC 479. Past studies have shown that after experiencing environmental effects such as ram pressure stripping, the remaining gas becomes less efficient at forming stars, with an even stronger impact on pairs \citep{Kim2023}. If that is the case, galaxy mergers may drive gas inflow toward the center, triggering a central starburst—similar to what is observed in isolated mergers \citep{Kado-Fong2020, Zhang2020b, Gao2022}. VCC 479 has a starburst center, but its integrated SFR places it below the star formation main sequence followed by ordinary galaxies (Section \ref{SED-global}), plausibly due to its gas-poor nature.

VCC 479 serves as a compelling case of a gas-poor dwarf-dwarf merger occurring in a dense environment.
The dwarf-dwarf mergers that occur in the cluster environment have already been reported \citep{Paudel2017,Zhang2020b}, despite the generally high velocity dispersion of member galaxies in clusters. VCC 479 is located on cluster outskirts, where the velocity dispersion is lower compared to the cluster core \citep{Boselli2014}. The moderate velocity dispersion on the outskirts of galaxy clusters may increase the likelihood of dwarf-dwarf interactions and mergers. Additionally, hierarchical accretion of groups plays a significant role in shaping cluster populations \citep[e.g.,][]{Lisker2018}. VCC 479 may have been accreted into the Virgo cluster as part of a galaxy group consisting of VCC 479, VCC 318, VCC 393, and VCC 343, with a dispersion velocity of $\sim$ 80 km s$^{-1}$, as shown in Figure \ref{fig:AVID}. In that case, we have observed a merger of dwarf galaxies within a group of dwarf galaxies, similar to the recent findings reported by \citet{Paudel2024}.

\subsection{The future evolution of VCC 479}\label{sec:bluecore}
Red dEs (dwarf ellipticals) in clusters are widely believed to be mostly transformed from late-type dwarfs \citep{Boselli2008,Lisker2009}. Among these, dEs with blue cores are thought to represent transitional types and have been extensively studied \citep{De2003,Gu2006,Lisker2006a,Tully2008,Pak2014,Chung2019,Rey2023}. These galaxies exhibit blue central regions with irregularities caused by recent or ongoing star formation, while their outer regions typically display colors similar to, or slightly bluer than, the colors of normal quiescent dEs without blue cores \citep{Lisker2006a,Pak2014}. Consequently, blue-core dEs are characterized by significant positive color gradients in their radial color profiles. For example, in the Virgo cluster, blue-core dEs show color differences exceeding 0.2 mag between their central and outer regions, as demonstrated by \citet{Lisker2006a}.
The blue core dwarfs are found to be ubiquitous in various environments, including clusters \citep{De2003, Lisker2006a, Urich2017}, groups \citep{Tully2008, Pak2014}, and the field \citep{Gu2006, Rey2023}. Previous studies suggest that blue core dEs are the result of the morphological evolution of late-type dwarfs due to cluster environmental effects, such as ram pressure stripping \citep{Lisker2006a, Boselli2008}. Recently, \citet{Chung2019} found evidence of past mergers for two blue-core dEs, based on their disturbed internal velocity fields and the existence of kinematically decoupled cores. These findings support the scenario of dwarf-dwarf mergers in lower-density environments as a mechanism for the formation of blue-core dEs. This merger scenario is also supported by simulations of gas-rich dwarf mergers \citep{Bekki2008}, which suggest that dwarf-dwarf merging triggers central bursts of star formation as gas falls in, forming massive compact cores dominated by young stellar populations. These galaxies first appear as BCDs and later evolve into blue-core late-type dwarfs after the BCD phase.

The color profiles of VCC 479 (right panel of Figure \ref{fig:SFdistribution}) show a pronounced positive gradient within the inner 10 arcsec radius. The \textit{g}$-$\textit{i} color difference between the center and outer regions is approximately 0.4 mag (right panel of Figure \ref{fig:SFdistribution}). Therefore, VCC 479 is a typical blue-core dwarf with central starburst sitting on a quiescent disk. Its formation is primarily dominated by a gas poor major merger that occurred in the past several hundred million years (Sections \ref{sec:shell} and \ref{SED-resolve}). VCC 479 provides the first direct and compelling evidence that dwarf mergers can lead to the formation of blue-core dwarfs.
We note that some blue-core dwarfs show an unusual enhancement of the oxygen abundance \citep[e.g.,][]{Peeples2008}. This does not seem to be the case for VCC 479 (Section \ref{sec:SDSS}), which is likely due to the central concentration of substantial HI gas (Section \ref{sec:HI-distribution}).

\begin{figure}
    \centering
    \includegraphics[width=\columnwidth]{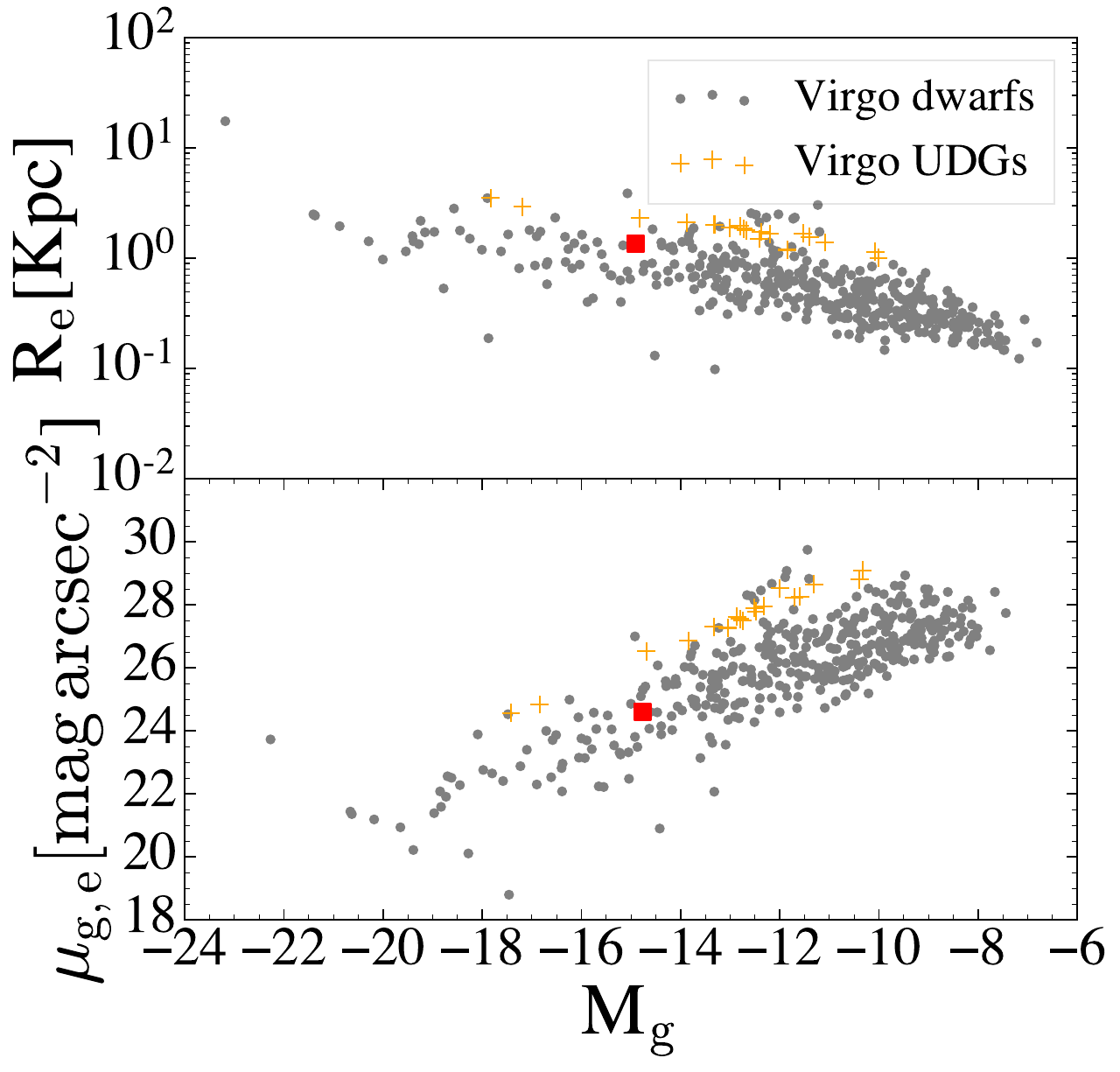}
    \caption{R$_e$-M$_g$ and $\mu_e$-M$_g$ scaling relations. The gray dots denote the dwarfs in the Virgo core region \citep{Ferrarese2020}, and the orange crosses denote the Virgo UDGs by \citet{Roediger2017}. VCC 479 is represented by a red square.}
    \label{fig:sr}
\end{figure}

The outer disk of VCC 479 shares color and structural characteristics with dEs. It shows no evidence of ongoing star formation (Section \ref{SED-global}). The shaded blue region in the right panel of Figure \ref{fig:SFdistribution} represents the \textit{g}$-$\textit{i} color range of normal dEs of comparable absolute magnitude, derived from the color-magnitude relation reported by \citet{Roediger2017}, based on NGVS data for dEs in the Virgo cluster's core. The colors of the quiescent disk of VCC 479 are slightly bluer than the colors of typical dEs in Virgo, which is expected given its location on the outskirts of the Virgo cluster \citep{Lisker2008}. This bluer outer color suggests a younger stellar population in the galaxy's main body, due to a later quenching than ordinary dEs (Section \ref{SED-resolve}). The outer exponential disk, with Sersi\'c index \textit{n} = 1, is consistent with the n-M$_{\rm B}$ relation for dEs in \citep[][]{Graham2003}. We show in Figure \ref{fig:sr} the R$_e$-M$_g$ and $\mu_e$-M$_g$ relation for dwarfs in the Virgo core region (central 0.3$\times$0.3 Mpc$^2$) by \citet{Ferrarese2020}. VCC 479 aligns well with these relations, indicating its structural properties are typical of dwarfs in the Virgo core. By contrast, BCDs generally fall significantly below these relations.

The stellar component of the quiescent disk is expected to relax to a dE-like smooth distribution in the next $\sim$ 1 Gyr or so \citep[][]{Paudel2017,Sazonova2021}, whereas the atomic gas (HI plus heavier elements) consumption timescale by star formation as observed now is over 5 Gyr, so VCC 479 is expected to evolve into a blue-core dE. This suggests that a fraction of the early-type dwarfs in the clusters have probably been shaped by dwarf–dwarf mergers \citep[e.g.,][]{Paudel2017,Chung2019}.

\section{Summary and conclusion} \label{sec:sum}
We have investigated the stellar and HI gas properties of a dwarf merger in detail using multifrequency observations, including VLA and FAST HI observations, NGVS images, VESTIGE narrow-band H$\alpha$ images, and an SDSS spectrum (Section \ref{sec:data}). VCC 479 is located on the outskirts of the Virgo cluster, and the location in the phase-space diagram indicates it is very likely a recent infaller of the Virgo cluster (Section \ref{sec:environ}).

For the stellar component, the symmetric shells along the major axis, visible in deep NGVS images, provide compelling evidence for a recent major merger (Section \ref{sec:shell}). Similar colors of the symmetrical shells and the stellar main body (Figure \ref{fig:slit-profile}) suggest that the two progenitor galaxies have similar stellar populations and thus likely have similar morphological types. In addition to the shells, the main characteristics of VCC 479 can be summarized as a central starburst embedded within a quiescent disk (Figure \ref{fig:SFdistribution} in Section \ref{SED-global}). The quenching of the outer disk is attributed to past environmental stripping of gas, whereas the centralized starburst is likely driven by gas inflow triggered by the merger (Section \ref{sec:merger}). The global SFR is lower than the star-forming main-sequence of ordinary galaxies. Using resolved SED fitting (Section \ref{sec:SED-extrac} and Section \ref{sec:SED-method}), we constrain the onset of the central starburst to approximately 600$^{+300}_{-140}$ Myr ago, and the quenching of the outer disk to around 900$^{+630}_{-290}$ Myr ago (Figure \ref{fig:SFH} in Section \ref{sec:SED-result}). These timescales are consistent with results from SDSS single-fibre spectral fitting (Section \ref{sec:SDSS}). This combination of a central starburst and a red, quiescent outer disk produces a significant color gradient (right panel of Figure \ref{fig:SFdistribution}), classifying VCC479 as a blue-core dwarf galaxy. This provides the first direct evidence that dwarf galaxy mergers can lead to the formation of blue-core dwarf galaxies.

For the HI gas, VCC 479 exhibits a high degree of HI deficiency and an unusually strong central concentration of HI, with negligible extended HI gas envelope (Figure \ref{fig:HI-map} in Section \ref{sec:HI-distribution}). Given that the HI gas velocity dispersion appears largely unaffected by the central starburst, the weak HI velocity field—misaligned with the photometric major axis—may reflect merger-driven HI gas inflow ( Section \ref{sec:HI-kin}).

We performed numerical simulations of dwarf–dwarf major mergers with varying gas fractions (Section \ref{sec:simulation}) to broadly reproduce the main characteristics of VCC 479, including the prominent symmetrical stellar shells, the centrally concentrated gas distribution, and the relatively modest stellar mass contribution from the merger-induced starburst.

The progenitors of VCC 479 may have lost most of their gas as a result of environmental effects prior to the final coalescence of the merger (Section \ref{sec:gaspoor}). The mass of newly formed stars in the starburst after the coalescence is approximately $2.4\times10^6\rm M_\odot$, accounting for only 2.9$\pm0.5$\% of the stellar mass (Section \ref{sec:SED-result}). The color and structural properties of VCC 479's outer disk align with those of dEs (Section \ref{sec:bluecore}). VCC 479 exhibits transitional properties between late-type and early-type dwarf galaxies and is expected to evolve into a blue-core dE once its quiescent stellar disk relaxes into a smooth configuration in the next $\sim$ 1 Gyr.

As the first case of gas-bearing major merger of dwarfs significantly influenced by environment, VCC 479 provides invaluable insights into the evolution of dwarf galaxies under the combined influence of environmental quenching and major mergers. Previous studies indicate that the HI gas distribution of interacting pairs of dwarf galaxies is more extended than that of isolated dwarfs, making them more susceptible to environmental stripping. This previous speculation aligns with the high HI deficiency of VCC 479. The substantial gas loss before final coalescence likely leads to less prominent merger-triggered star formation compared to that in isolated mergers. VCC 479 provides an unprecedented compelling case of coalesced major merger of late-type dwarf galaxies in a relatively dense environment.

\begin{acknowledgements}
We sincerely appreciate the anonymous referee's highly positive and encouraging report on our work. We acknowledge support from the National Key Re- search and Development Program of China (grant No. 2023YFA1608100), and from the NSFC (grant Nos. 12122303, 11973039, 12173079). This work is also supported by the China Manned Space Project (grant Nos. CMS- CSST-2021-B02 and CMS-CSST-2021-A07, CMS-CSST-2021-A06). We also acknowledge support from the CAS Pioneer Hundred Talents Program, the Strategic Priority Research Program of Chinese Academy of Sciences (grant No. XDB 41000000), and the Cyrus Chun Ying Tang Foundations. This work made use of the data from FAST (Five-hundred-meter Aperture Spherical radio Telescope)(https://cstr.cn/31116.02.FAST). FAST is a Chinese national mega-science facility, operated by National Astronomical Observatories, Chinese Academy of Sciences. We are grateful to Professor Jing Wang for her guidance in the design of the FAST observation strategy, and to her student Mr. Yang Dong for helpful suggestions during the observation process. RS acknowledges financial support from FONDECYT Regular 2023 project No.\ 1230441 and also gratefully acknowledges financial support from ANID -- MILENIO NCN2024\_112.
\end{acknowledgements}

\end{document}